\documentclass[10pt,lettersize,journal]{IEEEtran}
\usepackage{amsmath,amsfonts}
\usepackage{algorithmic}
\usepackage{algorithm}
\usepackage{array}
\usepackage[caption=false,font=normalsize,labelfont=sf,textfont=sf]{subfig}
\usepackage{textcomp}
\usepackage{stfloats}
\usepackage{url}
\usepackage{verbatim}
\usepackage{graphicx}
\usepackage{cite}
\usepackage{amsthm}
\usepackage{float,multirow,enumitem,multicol,multirow,amsfonts, subfig}
\newcommand{\bY}{\mathbf{Y}}
\newcommand{\by}{\mathbf{y}}
\newcommand{\boldE}{\textsf{E}}
\newcommand{\boldP}{\textsf{P}}
\newcommand{\bVar}{\text{var}}
\newcommand{\bE}{\textsf{E}}

\usepackage[x11names]{xcolor}

\newtheorem{theorem}{Theorem}[]

\newtheorem{lemma}[]{Lemma}

\newtheorem{condition}{Condition}
\graphicspath{{../}}

\allowdisplaybreaks 
\begin{document}

\title{Sequential Change-point Detection for High-dimensional and non-Euclidean Data}

\author{Lynna Chu and Hao Chen
\thanks{The work was supported in part by the NSF awards DMS-1513653 and DMS-1848579. }
\thanks{Lynna Chu is with the Department of Statistics, Iowa State University (email: lchu@iastate.edu).}
\thanks{Hao Chen is with the Department of Statistics, University of California, Davis (email: hxchen@ucdavis.edu).}
\thanks{This article has supplementary downloadable material available at https://doi.org/10.1109/TSP.2022.3205763, provided by the authors.}}
\markboth{}%
{Shell \MakeLowercase{\textit{et al.}}: A Sample Article Using IEEEtran.cls for IEEE Journals}


\maketitle

\begin{abstract}
In many applications, it is often of practical and scientific interest to detect anomaly events  in a streaming sequence of high-dimensional or non-Euclidean observations. We study a non-parametric framework that utilizes nearest neighbor information among the observations to detect changes in an online setting. It can be applied to data in arbitrary dimension and non-Euclidean data as long as a similarity measure on the sample space can be defined. We consider new test statistics under this framework that can detect anomaly events more effectively than the existing test while keeping the false discovery rate controlled at a fixed level. Analytic formulas approximating the average run lengths of the new approaches are derived to make them fast applicable to modern datasets. Simulation studies are provided to support theoretical results. The proposed approach is illustrated with an analysis of the NYC taxi dataset. 
\end{abstract}


\section{Introduction}

\IEEEPARstart{S}{equential} change-point detection aims to detect abrupt anomalies, observations that deviate from regular behavior, in a streaming sequence of observations as quickly as possible, while controlling the number of false alarms. 
In many modern applications, the sequence of observations may consist of high-dimensional observations (where the dimensional of each observation is larger than the sample size) or non-Euclidean objects (for example, sequences of networks or images). Examples include fraud detection involving large amounts of customer transactions \cite{bolton2001unsupervised}; disease surveillance or medical monitoring using sequences of images or multiple diagnostic measures \cite{dehning2020inferring}, \cite{pervaiz2012flubreaks}; and consumer-based data streams such as wearable and smart device monitoring, internet of things sensors, network security systems, cybersecurity, and other web applications \cite{zhang2013medmon, xie2011anomaly, tartakovsky2006novel}. In all these examples, the goal is to detect a change (if a change is present) as quickly as possible once it occurs, while also limiting the risk of false discovery. 

The sequential change-point setting can be formulated as follows: let the observation at $t$  be denoted as $\bY_t$, $t =1, 2, \hdots, n, \hdots$. Here, $t$ could be the time index or some other meaningful indices, $\bY_t$ could be a vector, image, or network, and $n$ is the index for the observation currently being observed. When there is no change-point, $\bY_t$'s are identically distributed from an unknown distribution, denoted as $F_0$. If there is a change-point at $\tau$, the observation after $\tau$ are from a different (unknown) distribution: 
\[ \bY_t \sim F_0,\ t = 1, \hdots, \tau - 1,\]
\[ \bY_t \sim F_1,\ t = \tau, \tau + 1, \hdots, \]
where $F_0$ and $F_1$ are two different probability measures. 

This formulation is very general. $F_0$ and $F_1$ are unknown and not specified.  We do not impose any constraints on how the change happens here. For example, if the observation is a high-dimensional vector, the change may occur in a subset of (unknown) data streams and the subset may be of size one.

Our aim is to construct a stopping rule (denoted as $T$) that will alert us to a change as quickly as possible, once it occurs, but keep the number of false discoveries at a fixed level. Moreover, in keeping with the spirit of the problem formulation, the stopping rule should be able to handle modern data types and should be relatively fast and easy to implement. To be precise, this means we would like to define $T$ such that the detection delay, $E_\tau(T-\tau|T>\tau)$, is small, subject to a fixed average run length (formally $E_\infty(T) \ge c$, where $c$ is a pre-specified large value) without making assumptions on the underlying sequence of observations. Here $E_\tau$ denotes the expectation under the hypothesis that the true change-point happens at $\tau$ and $E_\infty$ denotes the expectation under the hypothesis of no change. 


\subsection{Related Works} 
When the data is univariate (or scalar), the sequential change-point detection has been studied extensively (see \cite{siegmund1985sequential} and \cite{tartakovsky2014sequential}  for a review). For low-dimensional data, likelihood based methods have been explored which require knowledge or parameter estimates of the probability density functions (see for example: \cite{page1954continuous, lorden1971procedures, pollak1991sequential, lai1995sequential}). For high-dimensional data, the available methods are somewhat limited and may impose strict assumptions on the data. As an example, many existing methods have the assumption that the different data streams are independent \cite{tartakovsky2008asymptotically, mei2010efficient, xie2013sequential, wang2015large, chan2017optimal}. Other works allow for more flexible application; for example kernel-based methods \cite{desobry2005online} and a modified sliding window algorithm \cite{keriven2020newma}. Computationally efficient methods have also been proposed which combine summary statistics based on geometric entropy minimization (GEM) with the cumulative sum (CUSUM) algorithm \cite{yilmaz2017online, kurt2020real}.  For network data, \cite{zambon2018concept} proposed a sequential approach that acts by embedding each graph into a vector domain, where a conventional multivariate change-point detection procedure can be then applied. In general, for non-parametric methods applicable to high-dimensional and non-Euclidean data, theoretical analysis establishing false discovery control is very difficult to carry out. 

Recently, \cite{chen2019sequential} proposed a new non-parametric framework that utilizes nearest neighbor information to detect changes in an online setting. They also provided a general, analytical formula for false discovery control. This method can be applied to data in arbitrary dimensions (with no assumption that the different data streams are independent) and to non-Euclidean data. The author proposed to use the following stopping rule: 
\begin{equation}
T_Z(b_Z)  =  \inf \left\{n-N_0: \max_{n - n_1 \le t \le n-n_0}Z_{L|\by}(t,n) > b_Z \right \},
\label{eq:TZ} 
\end{equation} 
where $n_0, n_1$, and $L$ are pre-specified values, $N_0$ is the number of historical observations with no change-point, $n > N_0$, and $Z_{L|\by}(t,n)$ is a two-sample test statistic that tests whether $\{\bY_{n-L+1}, \dots, \bY_{t}\}$ and $\{\bY_{t+1}, \dots, \bY_{n}\}$ are from the same distribution. We refer to $Z_{L|\by}(t,n)$ as the \textit{edge-count two-sample test based on $k$-nearest neighbor} ($k$-NN) in the following for simplicity. For more details of this test, please see Section \ref{sec:newtests}. The author also provided an analytic formula to compute $b_Z$ such that the average run length $\bE_\infty(T_Z(b_Z))$ is controlled at a pre-determined value. Simulation studies show that this method beats likelihood-based methods when the dimension is high.  

Despite these nice properties, we find that the edge-count two-sample test on $k$-NN can have low power for some common types of changes when dimension is moderate to high, causing the stopping rule (\ref{eq:TZ}) to behave unexpectedly. To illustrate, consider a simple scenario where data are from a $d$-dimensional Gaussian distribution and there is a change at $\tau=201$: 
$$\bY_1, \hdots, \bY_{200} \overset{iid}{\sim}  \mathcal{N}_d(\mathbf{0}, \Sigma), \quad \bY_{201}, \hdots \overset{iid}{\sim} 
\mathcal{N}_d(\mathbf{\mu}, \sigma^2\Sigma)$$ with $\Sigma(i,j) = 0.3^{|i-j|}$. We consider two types of changes: 
\begin{itemize}
	\item Scenario 1 (only mean differs):  $||\mu||_2 = \Delta $. 
	\item Scenario 2 (both mean and variance differ): $||\mu||_2 = \Delta$ and $\sigma$. 
\end{itemize}

Tables \ref{table1} presents the performance of $T_Z(b_Z)$ for both scenarios based on 1,000 simulation runs. Here $k=5$, $L=200$, $n_0 = 25$, and $n_1 = 175$. The table reports the fraction of trials (out of 1,000) to successfully detect the change within 30 (or 50) observations after the change occurs. The average detection delay (EDD) is estimated as the average elapsed time between when the change occurs ($\tau = 201$) and when $T_Z(b_Z)$ detects a change. False alarms are not counted here. 
In each scenario, the threshold $b_Z$ is computed by formulas given in \cite{chen2019sequential} such that $\boldE_{\infty}(T_Z(b_Z)) = 2000$.  

\begin{table}[!t]
\caption{Performance of $T_Z(b_Z)$ for Scenarios 1 and 2, $\Delta = 2.6$, $\sigma = 0.78$, $d=100$}
\centering
 \begin{tabular}{|c |c c |} 
 \hline
  & Scenario 1 & Scenario 2  \\ [0.5ex] 
 \hline
 $<30$ & 0.15 & 0.06 \\ 
 $<50$ & 0.67 & 0.48\\
 \hline
 \multirow{2}{*}{EDD} & 44.29 & 50.21 \\
 & $\pm~14.17$ & $\pm~13.96$ \\
 \hline
 \end{tabular}
  \label{table1}
\end{table}

In Table \ref{table1}, we see that the method performs worse in Scenario 2 than Scenario 1 under two different comparison criteria: the fraction of trials that can be detected given a fixed time is smaller under Scenario 2 and the average detection delay (EDD) is larger under Scenario 2. 
However, common sense tells us that the additional change in variance should make the two distributions more different and  the change in Scenario 2 easier to detect. This phenomenon is due to the curse-of-dimensionality (a more detailed explanation can be found in Section \ref{sec:limitations}) and results in the existing method having diminished power to detect general changes. 

\subsection{Our contributions} \label{sec:contribution}

To address the problem of the stopping rule $T_Z(b_Z)$, we propose three new stopping rules:
\begin{align} 
& T_S(b_S) =  \inf \left\{n-N_0: \max_{n - n_1 \le t \le n-n_0}S_{L|\by}(t,n) > b_S \right \},  \label{eq:T_S}\\
&  T_{W}(b_W) = \inf \left\{n-N_0: \max_{n - n_1 \le t \le n-n_0} W_{L|\by}(t,n) > b_W \right \},  \label{eq:T_W}  \\
& T_M(b_M) =  \inf \left\{n-N_0: \max_{n - n_1 \le t \le n-n_0}M_{L|\by}(t,n) > b_M \right \}.  \label{eq:T_M} 
\end{align} 

The definitions of $S_{L|\by}(t,n)$, $W_{L|\by}(t,n)$, and $M_{L|\by}(t,n)$ are provided in Sections \ref{sec:Generalized}, \ref{sec:Weighted} and \ref{sec:Maxtype}, respectively. Under the same setup detailed in Table \ref{table1}, Table \ref{table2} shows that these new stopping rules are more successful in detecting the change quickly after it has occurred. They also have shorter detection delays than $T_Z(b_Z)$ under both above scenarios and all have shorter detection delays in Scenario 2 than that in Scenario 1. Further comparisons between the stopping rules can be found in Section \ref{sec:Power}; these demonstrate that the new stopping rule have improved power and detection delay compared to existing methods over a range of general scenarios. 
 \begin{table}[H]
\caption{Performance of new stopping rules under Scenario 1 (top) and Scenario 2 (bottom), $\Delta = 2.6$, $\sigma = 0.78$, $d=100$}
\centering
 \begin{tabular}{|c |c c c c|} 
 \hline
  & $Z$ & $W$ & $S$ & $M$ \\ 
 \hline
 $<30$ & 0.15 & 0.51  & 0.41 & 0.49\\ 
 $<50$ & 0.67 & 0.91 & 0.86 &  0.89\\
 \hline
\multirow{2}{*}{EDD} & 44.29 & 32.27 & 35.14 & $32.61$  \\
		& $\pm 14.17$ &  $ \pm 12.90$ & $\pm 15.48$ & $\pm 13.62$\\
 \hline
 \hline
 \end{tabular}
  \label{table2}
 \centering
 \begin{tabular}{|c |c c c c|} 
  
 & $Z$ & $W$ & $S$ & $M$ \\ 
 \hline
 $<30$ & 0.06 & 0.85  & 0.81 & 0.82\\ 
 $<50$ & 0.48 & 0.98 & 0.98 & 0.98 \\
 \hline
\multirow{2}{*}{EDD} & 50.21 & 23.21 & 24.16 & $23.74$ \\
		& $\pm~13.96$ & $\pm~8.01$ & $\pm~8.48$ & $ \pm~7.82$\\
 \hline
 \end{tabular}
\end{table}

To construct these new stopping rules, we propose new two-sample tests on $k$-NN.  Specifically, we extend the generalized ($S$) /weighted ($W$) /max-type ($M$) edge-count test defined on an undirected similarity graph \cite{chen2017new}, \cite{chu2019asymptotic} to the directed $k$-NN graph. The generalized edge-count and max-type tests on $k$-NN are well defined except for a particular construction of a $k$-NN graph (see Theorem \ref{THM:INVERTIBLE}). The detailed definitions of these stopping rules are given in Section \ref{sec:newtests}.

To make the new stopping rules useful for real-data applications, we provide analytical formulas to compute the stopping thresholds, $b_S, b_W$, and $b_M$, such that the average run length (ARL) for each new stopping rule is controlled at a pre-determined value.  This involves studying how the $k$-NN graph updates and obtaining expressions that explicitly characterize these dynamics, which lead to the development of new theoretical treatments. Specifically, for all the stopping rules, more accurate expressions of the graph updates are derived and the techniques used are an improvement over the approach utilized in \cite{chen2019sequential}.  
We demonstrate that the analytical formulas for the stopping thresholds are reasonable to use and we further improve upon their accuracy for finite sample sizes by implementing a skewness correction technique on the thresholds. 

In general, each test statistic has its own niche where it dominates. When interested in general change (for example, both mean and variance change), the stopping rules $T_S(b_S)$ and $T_M(b_M)$ are recommended. The stopping rule $T_M(b_M)$ has an advantage over $T_S(b_S)$ in that we can obtain more accurate analytical expression for the ARL for false discovery control. If the change of interest is in mean only, the stopping rule based on $T_W(b_W)$ is recommended. See Section \ref{sec:Power} for a comparison of their performance. In this paper, the types of changes we explore are confined to mean and/or variance change. However, the approach can be used to detect other changes in distribution, such as changes in covariance. For illustration, Table \ref{table_cov} shows the performance of the stopping rules under covariance change only. The data are again generated from $d $-dimensional Gaussian with $d = 100$,  $k=5$, $L=200$, $n_0 = 25$, $n_1 = 175$, and there is change at $\tau=201$: 
$$\bY_1, \hdots, \bY_{200} \overset{iid}{\sim}  \mathcal{N}(\mathbf{0}, \Sigma_0), \quad \bY_{201}, \hdots \overset{iid}{\sim} 
\mathcal{N}(\mathbf{0}, \Sigma_1),$$ with $\Sigma_0(i,j) = 0.3^{|i-j|}$ and $\Sigma_1(i,j) = 0.68^{|i-j|}$.

\begin{table}[H]
\caption{Performance of stopping rules under covariance change. }
\centering
 \begin{tabular}{|c |c c c c |} 
 \hline
  & Z  & W & S & M  \\ 
 \hline
 $<30$ & 0.026 & 0.263 & 0.336 & 0.334 \\ 
 $<50$ & 0.069 & 0.603 & 0.677 & 0.663\\
 \hline
 \multirow{2}{*}{EDD} & 92.17 & 65.65 & 43.90 & 45.27 \\
 & $\pm 27.69$ & $\pm 30.77$ & $\pm 25.85$ & $\pm 26.50$ \\
 \hline
 \end{tabular}
  \label{table_cov}
\end{table}

We see here again that the new stopping rules based on $W$, $S$, and $M$ perform better than the stopping rule based on $Z$. 

These new approaches are implemented in an \texttt{R} package \texttt{gStream}. 

The organization of the rest of the paper is as follows. Section \ref{sec:newtests} discusses the new stopping rules in details.  Section \ref{sec:ARL} studies the asymptotic properties of the proposed stopping rules and analytic ways to determine the thresholds.  The performance of the new methods are further explored in Section \ref{sec:Power} and the new methods are illustrated on a real data set in Section \ref{sec:Application}.

\section{New Tests} \label{sec:newtests}
The test statistics in the stopping rules (\ref{eq:T_S}) - (\ref{eq:T_M}) trace from the offline version of the problem studied in \cite{chu2019asymptotic} where the statistics were defined on an undirected similarity graph. For online detection, observations keep arriving and the similarity graph updates as a new observation arrives. Therefore one needs to understand the dynamics of the series of similarity graphs.  \cite{chen2019sequential} studied the directed nearest neighbor graphs as the dynamics of nearest neighbor graphs can be well understood.  In this work, we continue to use the directed nearest neighbor graphs to construct improved statistics.  The extension to other types of graphs is saved for future work.  


\subsection{Notation} \label{sec:notation}

First we define a random variable which indicates whether or not an observation $\bY_i$ is among the $r$th nearest neighbor to another observation $\bY_j$ among the observations in $n_L$.  
Specifically, for any $n > N_0$ and $i,j \in n_L \overset{\Delta}{=} \{ n- L + 1, \hdots, n \}$, we let
$A^{(r)}_{n_L,ij} = \text{I}$($\bY_j$ { is the } $r${th NN of } $\bY_i$  { among } $\bY_{n-L+1}, \hdots, \bY_n)$,
where $\text{I}(\cdot)$ is the indicator function that takes value $1$ if the event is true and $0$ otherwise. 
In terms of graph construction, each observation points to its $k$ nearest neighbors. For example, if $A^{(r)}_{n_L,ij} = 1$, then $\bY_j$ is the $r$th nearest neighbor of $\bY_i$ and there is a directed edge from $\bY_i$ pointing to $\bY_j$ (if $r \le k$).
We define $A^+_{n_L,ij} = \sum_{r=1}^k A^{(r)}_{n_L,ij}$ to be the indicator function that $\bY_j$ is one of the first $k$ NNs of $\bY_i$ among the observations in $n_L$. 
We use $\by_i$'s to denote the realizations of $\bY_i$'s and let $a^+_{n_L,ij} = \sum_{r=1}^k a^{(r)}_{n_L,ij}$ with  $a^{(r)}_{n_L,ij} = \text{I}( \by_j$ { is the } $r${th NN of } $\by_i$  { among }$\by_{n-L+1}, \hdots, \by_n)$.\
For any $n$, each $t \in \{n-L+1, \hdots, n \}$ divides the data sequence into two groups: one group being the observations before $t: \{\bY_{n-L+1}, \hdots, \bY_t \}$ (Group 1) and the other group being the observations after $t: \{\bY_{t+1}, \hdots, \bY_n \}$ (Group 2). 
Define,
\begin{align*} 
b_{0,ij}(t,n_L) = & \text{I}(n-L+1 \le i \le t \text{ and } t < j \le n \text{ or }\\
&  t < i \le n \text{ and } n-L+1 \le j \le t), \\
b_{1,ij}(t,n_L) = & \text{I}(n-L+1 \le i \le t \text{ and } n-L+1 \le j \le t), \\
b_{2,ij}(t,n_L) = & \text{I}(t<i \le n \text{ and } t < j \le n).
\end{align*}
Then $b_{0,ij}$ is the indicator function that $\bY_i$ and $\bY_j$ belong to different groups, $b_{1,ij}$ is the indicator function that $\bY_i$ and $\bY_j$ both belong to Group 1, and $b_{2,ij}$ is the indicator function that $\bY_i$ and $\bY_j$ both belong to Group 2.

We define our test statistics as follows: 
\begin{align*}
R_{0,L}(t,n) &= \sum_{i=n-L+1}^n \, \sum_{j=n-L+1}^n (A^+_{n_L,ij} + A^+_{n_L,ji})B_{0,ij}(t,n_L), \\ 
R_{1,L}(t,n) &= \sum_{i=n-L+1}^n \, \sum_{j=n-L=1}^n (A^+_{n_L,ij} + A^+_{n_L,ji})B_{1,ij}(t,n_L), \\ 
R_{2,L}(t,n) &= \sum_{i=n-L+1}^n \, \sum_{j=n-L=1}^n (A^+_{n_L,ij} + A^+_{n_L,ji})B_{2,ij}(t,n_L). 
\end{align*} 


$B_{0,ij}(t,n_L)$, $B_{1,ij}(t,n_L)$, and $B_{2,ij}(t,n_L)$ are the random variable versions of $b_{0,ij}(t,n_L)$, $b_{1,ij}(t,n_L)$, and $b_{2,ij}(t,n_L)$ such that the distribution of these random variables is defined to be the permutation distribution. The permutation distribution is the distribution induced by all $L!$ possible permutations of observation \textit{indices} among the observations in $n_L$. The null hypothesis and independence assumption imply that the observation indices are exchangeable, and therefore under the permutation distribution, the true (unknown) distribution of the observations remains unchanged when the null hypothesis is true.

\begin{figure}[!ht] 
    \centering
    \subfloat{
    \hspace{-5mm}

       \includegraphics[width=0.5\textwidth]{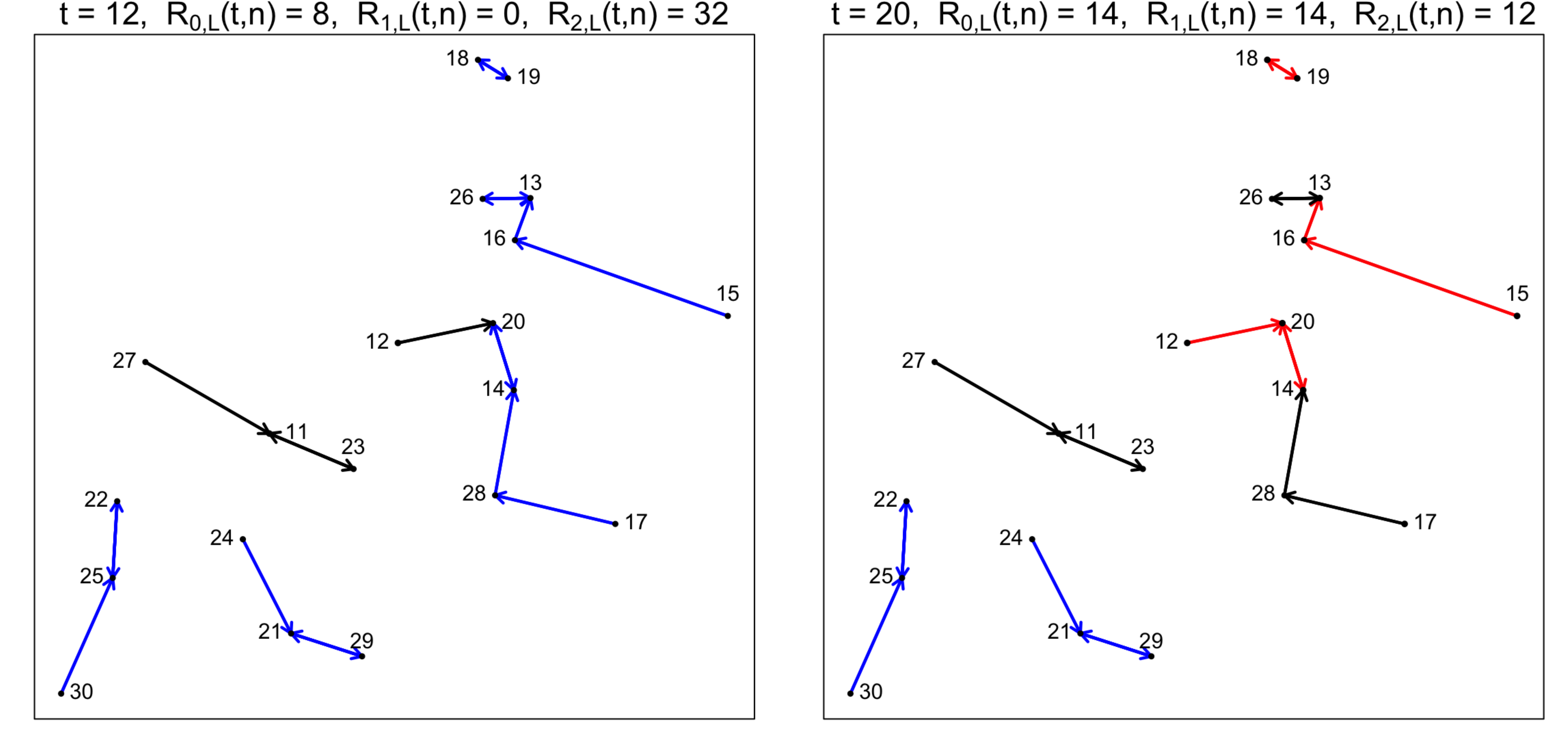}
       }
       \hspace{0mm}
          \subfloat{
             \hspace{-5mm}
       \includegraphics[width=0.5\textwidth]{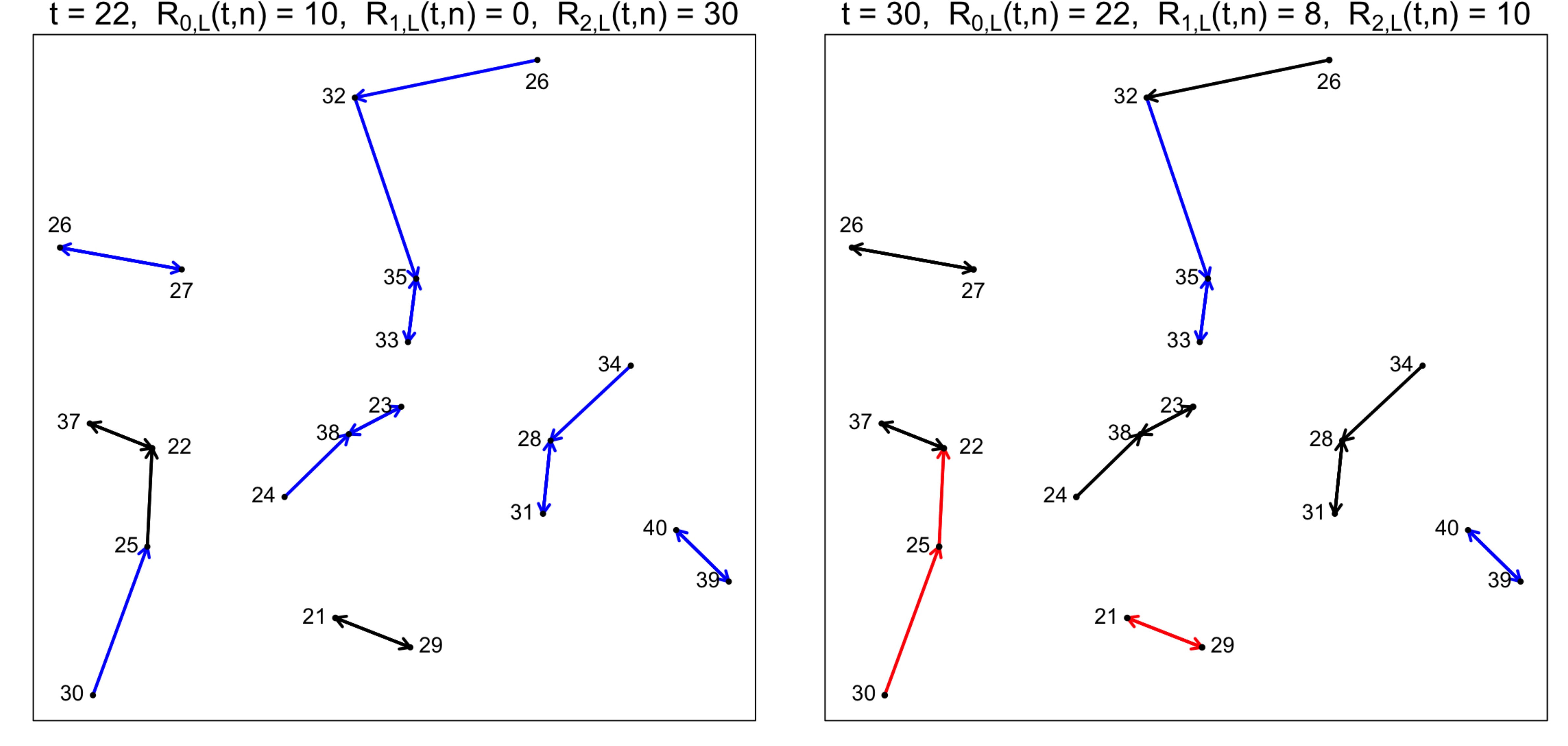}
       }
       \caption{The construction of graph-based test quantities $R_{0,L}(t,n)$ (black edges), $R_{1,L}(t,n)$ (red edges), and $R_{2,L}(t,n)$ (blue edges) for different values of $t$ on a directed $k$-NN graph. Each observation points to its $k$ nearest neighbors. Here $k=1$. In the first row, $n=30$ and the $k$-NN graph is constructed on the $L=20$ most recent observations: $y_{11}, y_{12}, \hdots, y_{30}$. In the second row, $n=40$ and the $k$-NN graph is constructed on the $L=20$ most recent observations: $y_{21}, y_{22}, \hdots, y_{40}$.}
       \label{fig:graph_quantities}
 \end{figure}

It is clear that $R_{0,L}(t,n)$ is twice  the number of edges in the $k$-NN graph connecting observations before $t$ and after $t$, $R_{1,L}(t,n)$ is twice the number of edges connecting observations prior to $t$, and $R_{2,L}(t,n)$ is twice the number of edges that connect observations after $t$. The notation of the graph-based test quantities emphasizes their dependency on the graph which is constructed on the $L$ most recent observations, with the most recent observation indexed at $n$. Figure \ref{fig:graph_quantities} illustrates these test statistics constructed for different times $t$ and $n$.  In the top row of Figure \ref{fig:graph_quantities}, $n = 30$ and we construct the graph on the $L = 20$ most recent observations: $y_{11}, y_{12}, \hdots, y_{30}$. The edge-counts are calculated for different values of $t \in \{11,\hdots, 30 \}$. $R_{0,L}(t,n)$ is twice the number of directed black edges, $R_{1,L}(t,n)$ is twice the number of directed red edges, and $R_{2,L}(t,n)$ is twice the number of directed blue edges. It is clear for a fixed $n$, the graph does not change:  each graph in the first row of Figure \ref{fig:graph_quantities} is the same. But as new observations continually arrive, the graph itself will update. For example when $n = 40$, we now construct the graph on observations $y_{21}, y_{22}, \hdots, y_{40}$ (see the second row of Figure \ref{fig:graph_quantities}). For $n=40$, the edge-counts are then calculated for different values of $t \in \{21, \hdots, 40\}$.

\subsection{Limitations of the method based on the edge-count test ($Z$)} \label{sec:limitations}

The stopping rule $T_Z(b_Z)$ (\ref{eq:TZ}) is based on the edge-count two-sample test statistic $Z_{L|y}(t,n)$ \cite{chen2019sequential}. To obtain $Z_{L|y}(t,n)$, first a directed $k$-NN graph is constructed based on a similarity measure (for example, Euclidean distance). Then the number of edges in the $k$-NN graph that connect observations before $t$ ($\bY_{n-L+1}, \hdots, \bY_t $) and after $t$ ($ \bY_{t+1}, \hdots, \bY_n $) is counted (we refer to this as the between-sample edge-count). Please see Figure \ref{fig:graph_quantities} for an illustration of how $R_{0,L}(t,n)$ is computed. 
To make $R_{0,L}(t,n)$ comparable across different $t$, we define its standardized version as
$ Z_{L|y}(t,n) = - \frac{R_{0,L}(t,n) - \boldE(R_{0,L}(t,n))}{\sqrt{\text{Var}(R_{0,L}(t,n))}}$.
Analytical expressions of $\boldE(R_{0,L}(t,n))$ and $\text{Var}(R_{0,L}(t,n))$ can be obtained under the permutation distribution, which is defined as all $L!$ possible rearrangements of the observation indices among $n_L$, and are omitted here for brevity.
A relatively low between-sample edge-count (or a large $Z_{L|y}(t,n)$) indicates the observations before and after $t$ are less mixed and this is evidence against the null hypothesis of no change. The intuition is that observations before $t$ tend to find their nearest neighbor among other observations before $t$ (and similarly for observations  after $t$), which implies a distributional difference between the two groups of observations. 

The rationale of a relatively small between sample edge-count ($R_{0,L}(t,n))$) holds particularly well under the scenario of mean change only and/or low-dimensional data. However, when the dimension of the data is moderate/high and the change in distribution is not only in mean, for example an additional variance change is present,  this rationale breaks down. This is because in the high-dimensional setting observations that are similar (i.e. from the same distribution) are not necessarily close in distance. Consider two distributions that differ in both mean and variance and the dimension of the data is $d=100$. The observations would be separated into two layers: observations with the smaller variance in the inner layer and observations with the larger variance in the outer layer. Since the volume of a $d$-dimensional space increases exponentially in $d$, for practical sample sizes it is not uncommon for observations in the outer layer to find themselves to be closer to observations in the inner layer compared to other observations in the outer layer. In this scenario the between-sample edge-count would be relatively large under the alternative, rendering $Z_{L|y}(t,n)$  ineffective. The new two-sample test statistics $S_{L|y}(t,n)$ and $M_{L|y}(t,n)$ address this curse-of-dimensionality problem. 

Moreover, even for mean change only the stopping rule based on $Z_{L|y}(t,n)$ can still suffer from increased detection delay.  This limitation of $Z_{L|y}(t,n)$ is due to a variance boosting problem under unequal sample sizes in the two-sample test setting. For more details on the variance boosting issue in the two-sample test setting see \cite{chen2018weighted}. In the sequential setting, this means that using $Z_{L|y}(t,n)$ can lead to increased detection delay since it may only be able to detect the change when it is near the middle of the sequence. 
To resolve this, we propose the weighted edge-count two-sample test $W_{L|y}(t,n)$. 

\subsection{The method based on the generalized edge-count test ($S$)} \label{sec:Generalized}

The generalized edge-count two-sample test at $t \in \{n-L+1, \hdots, n\}$ under $k$-NN can be defined as 
\begin{equation}
\label{eq:S}
S_{L|\by}(t,n)  =  \begin{pmatrix}
\bar{R}_{1,L}(t,n)) \\
\bar{R}_{2,L}(t,n)) \\
\end{pmatrix}^T (\mathbf{\Sigma_\by}(t,n))^{-1} \begin{pmatrix}
\bar{R}_{1,L}(t,n)  \\
\bar{R}_{2,L}(t,n)
\end{pmatrix}, 
\end{equation}
where $\bar{R}_{1,L}(t,n) = R_{1,L}(t,n) - \boldE(R_{1,L}(t,n)), \bar{R}_{2,L}(t,n) = R_{2,L}(t,n) - \boldE(R_{2,L}(t,n)) $, and  $\mathbf{\Sigma_\by}(t,n) = \text{Var}((R_{1,L}(t,n),R_{2,L}(t,n))^T|\by)$ such that $\boldE(\cdot)$ and $\text{Var}(\cdot)$ denote the expectation and variance taken under the permutation distribution. 

If a change-point $\tau > N_0$ occurs in the sequence, we would expect $S_{L|\by}(t,n)$ to be large when $n > \tau$ and $t$ close to $\tau$. The test statistic is defined in this way so that either direction of deviations in the number of within-sample edges from its null expectation would contribute to the test statistic. For example, under the location alternatives, we would expect both $R_{1,L}(t,n)$ and $R_{2,L}(t,n)$ to be larger than their null expectations, which would lead to a large $S_{L|\by}(t,n)$. Under the scale alternatives, the group with the smaller variance would have a within-edge count larger than its null expectation and the group with a larger variance would have a within-edge count smaller than its null expectation (due to the curse-of-dimensionality), which would also lead to a large $S_{L|\by}(t,n)$. Therefore, this test is powerful for both location and scale alternatives. 

Under the permutation distribution, the analytical expressions for $\boldE(R_{1,L}(t,n)|\by)$, $\boldE(R_{2,L}(t,n)|\by)$, and $\mathbf{\Sigma}_\by(t,n) = (\Sigma_{i,j}(t,n)|\by)_{i,j=1,2}$ can be calculated through combinatorial analysis. 
 Note that $\boldE(R_{1,L}(t,n)| \by) = \boldE(R_{1,L}(t,n))$ and $\boldE(R_{2,L}(x,n)| \by) = \boldE(R_{2,L}(t,n)).$ Let $x=t-(n-L)$. 
\begin{align*} 
& \boldE(R_{1,L}(x,n)) = \tfrac{2k x(x-1)}{(L-1)}, \\
& \boldE(R_{2,L}(t,n)) = \tfrac{2k(L-x)(L-x-1)}{(L-1)},  \\
& \Sigma_{\by,11}(x,n)   = \tfrac{4x(x-1)(L-x)}{(L-1)(L-2)(L-3)} ( (L-x-1)\times \\
& (k+ \tfrac{1}{L} \sum_{i,j \in n_L}a^+_{ij}a^+_{ji}) + (x-2)\tfrac{1}{L} \sum_{i,j \in n_L} a^+_{ji}a^+_{li}  - \tfrac{k^2x(L-3)}{(L-1)} ), \\
& \Sigma_{\by,22}(x,n)   = \tfrac{4x(L-x)(L-x-1)}{(L-1)(L-2)(L-3)} ( (x-1)(k+ \tfrac{1}{L} \sum_{i,j \in n_L}a^+_{ij}a^+_{ji}) \\
& \hspace{15mm} + (L-x-2)  \tfrac{1}{L} \sum_{i,j,l \in n_L} a^+_{ji}a^+_{li} - \tfrac{k^2(L-x)(L-3)}{(L-1)} ), \\
& \Sigma_{\by,12}(x,n)  = \Sigma_{\by,21}(x,n)  = \tfrac{4x(x-1)(L-x)(L-x-1)}{(L-1)(L-2)(L-3)}  ( \tfrac{k^2(L-3)}  {L-1} \\
& \hspace{15mm} + k + \tfrac{1}{L}\sum_{i,j \in n_L} a^+_{ij}a^+_{ji} - \tfrac{1}{L}\sum_{i,j \in n_L} a^+_{ji}a^+_{li} ).
\end{align*}

The generalized edge-count two-sample test statistic is well defined when $\Sigma_\by(t,n)$ is invertible.  

The following theorem ensures that the statistic is well defined except in very rare scenarios. These scenarios can be checked by calculating the node in-degree of each observation in the $k$-NN  graph. A node's in-degree is the number of other observations that find that node to be its nearest neighbors.  

\begin{theorem} \label{THM:INVERTIBLE} For $L\ge 5$, the generalized edge-count two-sample test statistic under $k$-NN is well defined except for when  all nodes have an in-degree of exactly $k$, i.e. $d_i = k$ $\forall i$, where  $d_i=\sum_{j\in n_L} a^{+}_{ji}$. 
\end{theorem} 

Based on Theorem \ref{THM:INVERTIBLE}, as long as each node in the graph does not have an in-degree of $k$ and the graph is constructed on at least $5$ observations, then $\Sigma_\by(t,n)$ is invertible and $S_{L|y}(t,n)$ is well-defined.
%
%
%
A proof of the above theorem is in the supplement.

This leads us to the stopping rule based on the generalized edge-count test under $k$-NN \eqref{eq:T_S}:
\begin{equation}
T_S(b_S) = \inf \left\{n - N_0:  \max_{n - n_1 \le t \le n-n_0} S_{L|\by}(t,n)  >b_S \right\}. \nonumber
\end{equation}

\subsection{The method based on the weighted edge-count test ($W$)} \label{sec:Weighted}
Following the same notations in Section \ref{sec:Generalized}, for each $t \in \{n-L+1, \hdots, n\}$, the weighted edge-count two-sample test statistic under $k$-NN can be defined as
$$R_{w,L}(t,n) = q(t,n) R_{1,L}(t,n) + p(t,n) R_{2,L}(t,n),$$
where $p(t,n) = \tfrac{t-(n-L)-1}{L-2}$ and $q(t,n) = 1- p(t,n)$. Since it is more difficult for the sample with a smaller sample size to form an edge within the same sample, $R_{1,L}(t,n)$ and $R_{2,L}(t,n)$ are weighted by the inverse of their corresponding sample sizes. The test statistic defined in this way resolves the variance boosting problem described in Section \ref{sec:limitations}. Relatively large values of $R_{w,L}(t,n)$ are evidence against the null hypothesis of no change. Let 
\begin{equation} \label{eq:Zw}
W_{L|\by}(t,n)  = \frac{R_{w,L}(t,n) - \boldE(R_{w,L}(t,n))}{\sqrt{\text{Var}(R_{w,L}(t,n)|\by)}}.
\end{equation} 

Under the permutation distribution, analytical formulas for $\boldE(R_{w,L}(t,n))$ and $\bVar(R_{w,L}(t,n)|\by)$ can be calculated based on $\bE(R_{1,L}(t,n))$, $\bE(R_{2,L}(t,n))$, and $\mathbf{\Sigma_\by}(t,n)$ provided in Section \ref{sec:Generalized}:
\begin{align*}
& \boldE(R_{w,L}(t,n)) = \tfrac{2kL(L-n+t-1)(n-t-1)}{(L-1)(L-2)}, \\
& \text{Var}(R_{w,L}(t,n)|\by)\\
& = \tfrac{4(L-n+t)(L-n+t-1)(n-t)(n-t-1)}{(L-1)(L-2)(L-3)} \times \\
&  ( k + \tfrac{\sum_{i,j \in n_L}a^+_{ij}a^+_{ji}}{L} - \tfrac{\sum_{i,j,l \in n_L} a^+_{ji}a^+_{li}}{L(L-2)} - \tfrac{k^2(L-3)}{(L-1)(L-2)}  ).
\end{align*} 

The variance of $R_{W|L}(t,n)$ is well defined if the inequality (\ref{cond_Rw}) holds.  Since $\sum_{i,j \in n_L} a^+_{ij}a^+_{ji} \ge 0$ by definition and $\frac{1}{L} \sum_{i,j,l \in n_L} a^+_{ji}a^+_{li}\le k(L-1)^2 + k^2$, we need: 
%
\begin{align}
\sum_{i,j \in n_L} a^+_{ij}a^+_{ji} & >\tfrac{k(L - 2k + Lk - 1)}{(L-1)(L-2)}= \tfrac{k}{(L-2)} + \tfrac{k^2}{(L-1)}. \label{cond_Rw}
\end{align}

The stopping rule based on the weighted edge-count test under $k$-NN is \eqref{eq:T_W}:
\begin{equation} 
T_W(b_W) =  \inf  \left\{ n- N_0:  \max_{n-n_1 \le t \le n- n_0} W_{L|\by}(t,n) > b_W \right \}. \nonumber
\end{equation}

\subsection{The method based on the max-type edge-count test ($M$)}   \label{sec:Maxtype}

We can define the max-type test statistic under $k$-NN based on the following lemma: 
\begin{lemma} \label{thm:gen_lemma} 
The generalized edge-count two-sample test under $k$-NN can be expressed as 
\[S_{L|\by}(t,n) = W_{L|\by}^2(t,n) + D_{L|\by}^2(t,n), \]
where $W_{L|\by}(t,n)$ is defined in (\ref{eq:Zw}), and 
\begin{equation} \label{eq:Zdiff}
D_{L|\by}(t,n) = \frac{R_{\text{diff},L}(t,n) - \boldE(R_{\text{diff},L}(t,n))}{\sqrt{\text{Var}(R_{\text{diff},L}(t,n)|\by)}} 
\end{equation}
with $R_{\text{diff},L}(t,n) = R_{1,L}(t,n) - R_{2,L}(t,n)$. 
\end{lemma}

The proof of this lemma is in the supplement. The analytical expressions for the expectation and variance of $R_{\text{diff},L}(t,n)$ under the permutation null are:
\begin{align*}
 \boldE(R_{\text{diff},L}(t,n)) & = 2k(L-2n+2t),\\
 \text{Var}(R_{\text{diff},L}(t,n)|\by) & = \tfrac{4(L-n+t)(n-t)}{(L-1)}(\tfrac{1}{L}  \sum_{ij \in n_L} a^+_{ji}a^+_{li}  - k^2).
\end{align*} 

The variance of $R_{\text{diff},L}(t,n)|\by$ is well-defined as long as $d_i \neq k \, \forall \, i$, in other words as long as each node does not have an in-degree of $k$. 

From the above lemma, $S_{L|\by}(t,n)$ is the sum of squares of two uncorrelated quantities (these two quantities are further asymptotically independent; details given in Section \ref{sec:ARL}). Here, $W_{L|\by}(t,n)$ is sensitive to location changes: when the change is in mean, $W_{L|\by}(t,n)$ tends to be large. On the other hand, $D_{L|\by}(t,n)$ is more sensitive to scale changes: when the change is in variance, $|D_{L|\by}(t,n)|$ tends to be large. The sign of $D_{L|\by}(t,n)$ depends on whether the distribution after the change has a larger spread or not. This leads to the following max-type edge-count two-sample test statistic under $k$-NN: 
\begin{equation} 
M_{L|\by}(t,n) = \max(|D_{L|\by}(t,n)|, W_{L|\by}(t,n)).
\end{equation}

When there is a change in location and/or scale, depending on the signal of interest, it is useful to consider an extended version of the max-type edge-count two-sample test: 
\begin{equation} 
M_{\xi,L|\by}(t,n) = \max(|D_{L|\by}(t,n)|,\xi W_{L|\by}(t,n)),
\end{equation}
where $\xi \ge 0$. Different choices of $\xi$ lead to different focuses of the alternatives. For example, if we are more interested in locational changes, we could choose a large $\xi$. On the other hand, setting $\xi$ to be small would favor detecting scale changes. When $\xi =1$, the test reduces to the plain max-type edge-count test. For more detailed discussion on how to select $\xi$, see Supplement H in \cite{chu2019asymptotic} (under the offline change-point detection setting, but similar arguments apply to the online setting).

The stopping rule based on the max-type edge-count test under $k$-NN  is as follows :
\begin{align} 
& T_{M_{\xi}}(b_{M_{\xi}}) =  \inf \left\{n - N_0:  \max_{n_1' \le t \le n_0'} M_{\xi,L|\by}(t,n)  >b_{M_\xi} \right\}, \label{eq:T_Mkappa} 
\end{align}
 where $n_1' = n - n_1$ and $n_0' = n - n_0'$.  This reduces to \eqref{eq:T_M} when $\xi = 1$. 

\section{Average run length} \label{sec:ARL}

Given the new stopping rules presented in Section \ref{sec:newtests}, we would like to determine the thresholds $b_S$, $b_W$, and $b_{M_\xi}$ in an analytic way such that the false discovery rate is controlled at a pre-specified value.  A common way to measure the false discovery rate under the online change-point detection is the average run length, i.e., the expected time to stop when there is no change-point, which we denote as $\boldE_\infty(T_S(b_S))$, $\boldE_\infty(T_W(b_W))$, and $\boldE_\infty(T_{M_\xi}(b_{M_\xi}))$. 

In the comparisons in Section \ref{sec:contribution} (Table \ref{table2}), the thresholds were chosen such that the average run lengths are 2,000 based on simulation runs.  This is doable when the underlying distribution of the sequence is known. However, in many applications, the distribution of the sequence is unknown. 
Furthermore, since new observations keep arriving, resampling based methods, such as permutation and bootstrap, are not appropriate here and even if they were, directly resampling could be very time consuming. Therefore, to make the method fast applicable, we seek to derive analytical expressions for the average run lengths. Given the non-parametric nature of the proposed method, we would not be able to get exact analytic formulas for the average run lengths under finite $L$.  In the following,  we first approach the problem asymptotically (Section \ref{sec:asym}), and then make adjustments for finite samples (Section \ref{sec:Finite}).   

\subsection{Asymptotic results} \label{sec:asym}

To derive analytical expressions for the average run length, we must study the asymptotic distribution of the stopping rules. Since these stopping rules are composed of the random fields $\{S_{L|\by}(t,n_L) \}$, $\{W_{L|\by}(t,n) \}$, and $\{M_{\xi,L|\by}(t,n)\}$, we study their asymptotic properties. To obtain the limiting distribution of these random fields, we only need to focus on $\{D_{L|\by}(t,n) \}$ and $\{W_{L|\by}(t,n) \}$. We show that the limiting distribution of the random fields  converge to independent two-dimensional Gaussian random fields (Theorem \ref{thm:gaussian}). This proof utilizes Stein's method  \cite{chen2005stein}.  To fully specify the Gaussian processes, we must derive the covariance functions of the new processes (Theorem \ref{thm:cov}). Since the $k$-NN graph updates each time new observations arrive, we must study the dynamics of the $k$-NN series for the new test statistics. While \cite{chen2019sequential} laid a framework for graph-based sequential detection, the techniques developed in \cite{chen2019sequential} were not directly applicable to the stopping rules proposed in this manuscript and needed to be adapted. Integration of techniques from \cite{chu2019asymptotic} were useful in this development. However, since the new stopping rules can cover more types of change, we found that a direct extension of these previous works were not sufficient in developing accurate analytical formulas approximating ARL. To push forward the theory, we carefully studied the dynamics of directed $k$-NN graphs and considered additional ways the graph can update. A more in-depth understanding of the graph updates led us to derive analytical expressions for these updates and incorporate them into our analytical formulas for the thresholds. A comparison of the improvement over a direct extension of previous works can be found in the supplement. Putting together Theorems \ref{thm:gaussian} and \ref{thm:cov} with results from \cite{siegmund1995using}, allow us to obtain analytical expressions for the average run lengths of the new stopping rules. 

In the following, we provide a sketch of the key steps to obtain each result. We defer readers to the supplement for a more technical treatment. 
The subsequent results are derived under the following condition: 
\begin{condition} \label{indeg_bounded} There is a positive constant $\mathbb{C}$, $1 \le \mathbb{C} < \infty$, depending only on $k$, such that 
\[\sup_{1 \le j \le n} \left( \sum_{i=1}^n A^+_{n,ij} \right) \le \mathbb{C}, \quad n \in \mathbb{N}. \]
\end{condition} 

In $k$-NN, each observation points to its first $k$ NNs, so the out-degree of each observation is $k$. However the in-degree of each observation can vary. This condition states that the in-degree of each observation is bounded. It is satisfied almost surely for multivariate data \cite{bickel1983sums},\cite{henze1988multivariate}. 

Let
\[D_{L}(t,n) = \frac{R_{\text{diff},L}(t,n) - \boldE(R_{\text{diff},L}(t,n))}{\sqrt{\text{Var}(R_{\text{diff},L}(t,n))}}, \]
\[W_{L}(t,n) = \frac{R_{w,L}(t,n) - \boldE(R_{w,L}(t,n))}{\sqrt{\text{Var}(R_{w,L}(t,n))}}, \]
which replaces the conditional variances of the test statistics with the unconditional variances. See the supplement for more details.

{Our proof that the limiting distributions converge to independent two-dimensional random fields depends on utilizing Stein's method \cite{chen2005stein}. The general idea of Stein's method is to show for a random variable $W$ and a standard normal random variable $Z$, that for some family of functions $h \in Lip(1)$, the following bound holds: 
 \[ \sup_{h \in Lip(1)} |\boldE h(W) - \boldE h(Z)| \le \delta, \] 
where $\delta$ depends on the structure of $W$. The specific form of Stein's method used here requires $A^+_{n_L,ij}$ to be locally dependent. However, 
even for different  $i$, $j$, $l$, $r$, $A^+_{n_L,ij}$ and $A^+_{n_L,lr}$ are dependent due to the constraint that $\sum_{j\in n_L} A^+_{n_L,ij} = k$ for all $i \in n_L.$
Following \cite{chen2019sequential}, we relax these dependencies by considering a similar set of Bernoulli random variables $\{ \tilde{A}^+_{n_L,ij}\}_{i,j \in n_L}$. We keep the following probabilities unchanged:
\begin{align*} 
\boldP(\tilde{A}^+_{n_L,ij} = 1) \quad & =\quad \boldP(A^+_{n_L,ij}=1), \\
\boldP(\tilde{A}^+_{n_L,ij} = 1, \tilde{A}^+_{n_L,ji} = 1) \quad & = \quad \boldP(A^+_{n_L,ij}=1, A^+_{n_L,ji}=1), \\
\boldP(\tilde{A}^+_{n_L,ji} = 1, \tilde{A}^+_{n_L,li} = 1) \quad & = \quad \boldP(A^+_{n_L,ji}=1, A^+_{n_L,li}=1),
\end{align*} 
but relax the other dependencies such that $\tilde{A}^+_{n_L,ij}$ is independent of $\{\tilde{A}^+_{n_L,il}, \tilde{A}^+_{n_L,li} \}_{l \neq j}$, and $\tilde{A}^+_{n_L,ij}$ and $\tilde{A}^+_{n_L,lr}$ are independent when $i$, $j$, $l$, $r$ are all different. Then $\tilde{A}^+_{n_L,ij}$ are only locally dependent and can be analyzed through the Stein's method \cite{chen2005stein}.  }

We are now ready to present the main results. 

\begin{theorem} \label{thm:gaussian} Under Condition \ref{indeg_bounded}, as $L \rightarrow \infty$, the finite dimensional distributions of $\{D_{L}([uL],[vL]): 0 < v -1 < u < v < \infty \} $ and $\{W_{L}([uL],[vL]): 0 < v -1 < u < v < \infty \} $ converge to independent two-dimensional Gaussian random fields, which we denote as  $\{D^\star(u,v): 0 < v -1 < u < v < \infty \} $ and $\{W^\star(u,v): 0 < v -1 < u < v < \infty \}$, respectively. Here $[x]$ denotes the largest integer smaller than or equal to $x$ for any real number $x$. 
\end{theorem}
The detailed proof for Theorem \ref{thm:gaussian} is in  the supplement.

Based on Theorem \ref{thm:gaussian}, we can approximate $\boldE_\infty(T_S(b_S))$, $\boldE_\infty(T_W(b_W))$, and $\boldE_\infty(T_{M_\xi}(b_{M_\xi}))$ 
by examining the asymptotic behavior of our stopping rules: 
\begin{align}
& T^\star_S(b_S)  = \inf \left \{ n-N_0: \max_{n_1' \le t \le n_0'} S^\star(\tfrac{t}{L},\tfrac{n}{L})  > b_S \right \}, \\
& T^\star_w(b_W)  = \inf \left \{ n-N_0: \max_{n_1' \le t \le n_0'} W^\star(\tfrac{t}{L},\tfrac{n}{L})  > b_W \right \}, \\
& T^\star_{M_\xi}(b_{M_\xi}) = \inf \left \{ n-N_0: \max_{n_1' \le t \le n_0'} M_{\xi}^\star(\tfrac{t}{L},\tfrac{n}{L})  > b_{M_\xi} \right \},
\end{align}
where $n_1' = n - n_1$ and $n_0' = n - n_0'$. 
Our approximations involve the function $\nu(x)$ defined as
$$\nu(x)  = 2x^{-2}\exp  \{ -2 \sum_{m=1}^\infty m^{-1}  \Phi \left( -\tfrac{1}{2}xm^{1/2} \right)  \}, x>0.$$ This function is closely related to the Laplace transform of the overshoot over the boundary of a random walk.  A simple approximation given in \cite{siegmund2007statistics} is sufficient for numerical purposes:
$$ \nu(x) \approx \frac{(2/x)(\Phi(x/2)- 0.5)}{(x/2)\Phi(x/2) + \phi(x/2)}, $$
where $\Phi(\cdot)$ is the cumulative distribution function of the standard normal distribution and $\phi(\cdot)$ the density function of the standard normal distribution.

According to \cite{siegmund1995using}, $T_S^\star(b_S)$, 
$T_W^\star(b_W)$, and $T_D^\star(b_D)$ are asymptotically exponentially distributed
$T_S^\star(b_S) \sim \exp(\lambda_S)$, $T_W^\star(b_W) \sim \exp(\lambda_W)$, $T_D^\star(b_D) \sim \exp(\lambda_D)$, 
 with means:
\begin{align}
 \boldE_{\infty}(T_S^\star(b_S)) & = \frac{1}{\lambda_S} \approx \frac{\pi \exp(b_S/2)}{c_0b_S H(c_0,h_1,h_2)}    \\
 \boldE_{\infty}(T_W^\star(b_W)) & = \frac{1}{\lambda_S}  \approx \frac{\sqrt{2\pi}\exp(b_W^2/2)}{c_1^2b_W G(c_2,g_{W,1},g_{W,2})}\\
 \boldE_{\infty}(T_D^\star(b_D)) & = \frac{1}{\lambda_D} \approx \frac{\sqrt{2\pi}\exp(b_D^2/2)}{2c_2^2b_D G(c_2,g_{D,1},g_{D,2})}  \\
 \boldE_{\infty}(T_{M_\xi}^\star(b_{M_\xi})) &  \approx 
 \end{align}
 \begin{align}
 & \begin{cases} \dfrac{\boldE_{\infty}(T^\star_D(b_{M_\xi})) \boldE_{\infty}(T^\star_W(b_{M_\xi}/\xi))}{ \boldE_{\infty}(T^\star_D(b_{M_\xi})) +  \boldE_{\infty}(T^\star_W(b_{M_\xi}/\xi))  } 
&  \text{ when } \xi > 0,  \vspace{2mm}\nonumber
\\\xi
\boldE_{\infty}(T_D^\star(b_{M_\xi})) &  \text{ when } \xi = 0,
\end{cases}
 \end{align}
with
\begin{align*}
&H(c,h_1,h_2) = \int_{0}^{2\pi} \int_{x_0}^{x_1} \left\{ h_{1}(x,\omega)h_{2}(x,\omega) \times \right. \\
& \hspace{20mm} \left. \nu( \sqrt{2 c\, h_{1}(x,\omega)}) \nu( \sqrt{2 c \,h_{2}(x,\omega)}) \right\} dx d\omega,\\
&G(c,g_1,g_2)  =  \int_{x_0}^{x_1} g_{1}(x)g_{2}(x) \nu(c \sqrt{2g_{1}(x)}) \nu(c \sqrt{2g_{2}(x)}) dx,
\end{align*} 
\begin{align*} 
g_{W,1}(x) &  = \tfrac{\partial_{-} \rho^{\star}_W(\delta_1,0)}{\partial \delta_1} \Bigr|_{\delta_1 = 0}  \equiv - \tfrac{\partial_{+} \rho^{\star}_W(\delta_1,0)}{\partial \delta_1} \Bigr|_{\delta_1 = 0}, \\
g_{D,1}(x) &  = \tfrac{\partial_{-} \rho^{\star}_D(\delta_1,0)}{\partial \delta_1} \Bigr|_{\delta_1 = 0}   \equiv - \tfrac{\partial_{+} \rho^{\star}_D(\delta_1,0)}{\partial \delta_1} \Bigr|_{\delta_1 = 0}, \\
g_{W,2}(x) & =  -\tfrac{\partial_{+} \rho^{\star}_W(\delta_2,0)}{\partial \delta_2} \Bigr|_{\delta_1 = 0}, \\
g_{D,2}(x) & =  -\tfrac{\partial_{+} \rho^{\star}_D(\delta_2,0)}{\partial \delta_2} \Bigr|_{\delta_1 = 0}, \\
h_1(x,\omega) & = g_{W,1}(x)\sin^2(\omega)+g_{D,1}(x)\cos^2(\omega), \\
h_2(x,\omega) & =  g_{W,2}(x)\sin^2(\omega)+g_{D,2}(x)\cos^2(\omega).
\end{align*}

We now have approximations of the average run lengths for the new stopping rules. The only remaining unspecified quantities are the directional partial derivatives of the covariance functions of the Gaussian random fields. Their analytical expressions are derived in the following theorem.

 \begin{theorem} \label{thm:cov} For two-dimensional fields $\{W^\star(u,v): 0 < v-1 < u < v < \infty\}$ and $\{D^\star(u,v): 0 < v-1 < u < v < \infty\}$, the directional partial derivatives are
\begin{align*}
g_{W,1}(x) & = \frac{1}{x(1-x)}, \\
g_{W,2}(x) & = \frac{x^2-x+1}{x(1-x)} - \frac{2 k\, p_{k+1,\infty}^{(k)}}{k + p_{k,\infty}}, \\
g_{D,1}(x) & = \frac{1}{2x(1-x)}, \\
g_{D,2}(x) & = \frac{10 q_{k,\infty} - 4k q_{k+1,\infty}^{(k)} - (6k^2-10k)}{2 (q_{k,\infty}-k^2+k)} - \frac{1}{2x(1-x)}, 
\end{align*} 
where 
\begin{align*}
& p_{k+1,\infty}^{(k)} = \sum_{r=1}^k p_{\infty}(r,k+1),\quad q_{k+1,\infty}^{(k)} = \sum_{r=1}^k q_{\infty}(r,k+1).
\end{align*}
\end{theorem}
Here, $p_{k,\infty}$ is the limiting expected number of mutual NNs a node has in $k$-NN, $q_{k,\infty}$ is the limiting expected number of nodes that share a NN with another node in $k$-NN, $p_{\infty}(r,s)$ is the limiting expected number of mutual NNs shared between the $r$th and $s$th NNs, and similarly $q_{\infty}(r,s)$ is the limiting expected number of nodes shared between the $r$th and $s$th NNs. {Explicitly, 
$ p_{k,\infty}  = \sum_{r=1}^k \sum_{s=1}^k p_{\infty}(r,s),$ and $q_{k,\infty}  = \sum_{r=1}^k \sum_{s=1}^k q_{\infty}(r,s)$, with  $ p_\infty(r,s)  = \lim_{L \rightarrow \infty} \tfrac{1}{L} \sum_{i,j \in n_L} A^{(r)}_{n_L,ij} A^{(s)}_{n_L,ji}$ and $ q_\infty(r,s)  = \lim_{L \rightarrow \infty} \tfrac{1}{L} \sum_{i,j,l \in n_L, j \neq l} A^{(r)}_{n_L,ji} A^{(s)}_{n_L,li}$.}

To derive these partial derivatives, we studied the dynamics of the $k$-NN series as new observations are added. It turns out that a few key quantities are enough to characterize the dynamics in the asymptotic domain. The proof of this theorem is in the supplement. 


\subsection{Finite $L$} \label{sec:Finite} We now consider the practical scenario where $L$ is finite. Based on results in Section \ref{sec:asym}, $\boldE_\infty(T_S(b_S))$, $\boldE_\infty(T_W(b_W))$, and $\boldE_\infty(T_{M_\xi}(b_{M}))$ can be approximated by 
\begin{align}
 & \boldE_{\infty}(T_S(b_S)) \approx \frac{L\pi \exp(b_S/2)} {b_S^2 H_L(b_S,h_1,h_2)}  \label{eq:Ts}\\
& \boldE_{\infty}(T_W(b_W))  \approx \frac{L\sqrt{2\pi}\exp(b_W^2/2)}{ b_W^3 G_L(b_W,g_{W,1},g_{W,2})},  \label{eq:Tw}  
\end{align}
\begin{align}
 & \boldE_{\infty}(T_{M_\xi}(b_{M_\xi})) \nonumber  \\
 & \approx  \begin{cases} \dfrac{\boldE_{\infty}(T_D(b_{M_\xi})) \boldE_{\infty}(T_W(\tfrac{b_{M_\xi}}{\xi}))}{ \boldE_{\infty}(T_D(b_{M_\xi})) +  \boldE_{\infty}(T_W(\frac{b_{M_\xi}}{\xi}))  } &  \text{ when } \xi > 0, \nonumber \\
\boldE_{\infty}(T_D(b_{M_\xi})) &  \text{ when } \xi = 0,
\end{cases} \nonumber \\
& \text{ where }  \boldE_{\infty}(T_D(b_D)) \approx  \frac{L\sqrt{2\pi}\exp(b_D^2/2)}{2b_D^3 G_L(b_D,g_{D,1},g_{D,2})}, \label{eq:Tm}
\end{align}
with $H_L()$ and $G_L()$ are finite sample versions of $H()$ and $G()$, respectively. 
In practice, when $L$ is finite we use $g_{W,2}(L,x)$ and $g_{D,2}(L,x)$ in place of $g_{W,2}(x)$ and $g_{D,2}(x)$ in the above formulas, respectively, where 
\begin{align*} 
g_{W,2}(L,x) & = \tfrac{x^2-x+1}{x(1-x)} - \tfrac{2 k\, p_{k+1,L}^{(k)}}{k + p_{k,L}}, \\
g_{D,2}(L,x) &  =  \tfrac{10 q_{k,\infty} - 4k q_{k+1,L}^{(k)} - (6k^2-10k)}{2 (q_{k,L}-k^2+k)} - \tfrac{1}{2x(1-x)}.
\end{align*} 
Here, $p_{k,L}$, $p^{(k)}_{k+1,L}$, $q_{k,L}$, and $q^{(k)}_{k+1,L}$ are the finite sample versions of $p_{k,\infty}, p_{k+1,\infty}^{(k)}, q_{k,\infty}$, and $q_{k+1,\infty}^{(k)}$ and can be estimated in a data-driven way.

\subsection{Skewness correction} Analytical approximations provided in Section \ref{sec:Finite} become less precise for finite $L$ when $n_0$ is relatively small. This is mainly because the convergence of $W_L(t,n)$ and $D_L(t,n)$ to normal is slow if $(n-t)/L$ is close to $0$ or $1$. This problem becomes more severe when dimension is high. To improve upon the analytic approximations for finite sample sizes, we perform skewness correction. We adopt a skewness correction approach discussed in \cite{chen2015graph} that does the correction up to different extents based on the amount of skewness at each value of $t$. In particular, we provide better approximations to the marginal probabilities $\boldP(W^\star(u-x,w) \in b + du)$ and $\boldP(D^\star(u-x,w) \in b + du)$. Following the method based on cumulant-generating functions and change of measure (details refer to \cite{chen2015graph}), we can approximate the marginal probability by 
\[ \frac{1}{\sqrt{2\pi(1+\gamma\theta_b)}} \exp(-\theta_b - u \theta_b/b + \theta_b^2(1+\gamma \theta_b/3)/2), \]
where $\theta_b$ is chosen such that $\dot{\psi}(\theta_b) = b$. By a third Taylor approximation, we get $\theta_b \approx (-1+\sqrt{1+2\gamma_L(t,n) b})/\gamma_L(t,n)$, where $\gamma_L(t,n) := \boldE_P(Z_L(t,n)^3)$ and explicit expressions are derived using combinatorial analysis. 

Skewness corrected thresholds are only obtained for $W_{L|\by}(t,n)$ and $D_{L|\by}(t,n)$, but not $S_{L|\by}(t,n)$. This is because for $S_{L|\by}(t,n)$ the integrand can easily be non-finite and the approach depends heavily on extrapolation. Therefore, the stopping rule based on $M_{L|\by}(t,n)$ is often recommended because it can detect general changes but we can obtain more accurate stopping thresholds. 

\subsection{Checking accuracy of analytic formulas for the average run lengths}
Here, we check the accuracy of the analytic formulas for the average run lengths.  For all three new tests, we have analytic formulas based on asymptotic results (\ref{eq:Ts}), (\ref{eq:Tw}) and (\ref{eq:Tm}), and for the tests based on the weighted/max-type edge-count tests, we have analytic formulas after skewness correction, provided in the supplement.   We compare the empirical ARL obtained from these analytic formulas (with skewness correction with applicable) to those obtained from $1,000$ Monte Carlo simulations. The analytical thresholds are obtained so that the average run length is $2,000$. We generated data from three different settings: multivariate normal with $d =10$ (denoted by $C1$), multivariate $t_5$ with $d =100$ (denoted by $C2$), and multivariate log-normal with $d = 1000$ (denoted by $C3$). 

Results for different choices of $n_0$ are shown in Tables \ref{table:b_S_t5} - \ref{table:b_M_t5}. We set $n_1 = L-n_0$ and $k=5$. The asymptotic analytic results are denoted by `A1' and the skewness corrected approximations are denoted by `A2'.  We see in general that the empirical ARLs obtained from the asymptotic approximations (`A1') are not very close, illustrating the need for skewness correction here. After skewness correction, the empirical ARLs are much closer to $2,000$. It is clear that the accuracy of the skewness corrected approximations depends on $n_0$: in general, when $n_0 = 40$, the skewness corrected approximations do well across most dimensions. 

We also investigate how the analytical threshold approximations perform compare to the Monte Carlo thresholds in the presence of change. In Tables  \ref{table:b_S_t5_2}, \ref{table:b_W_t5_2}, and \ref{table:b_M_t5_2} we report the power, defined as the fraction of trials able to detect the change within 50 observations after the change occurs, and the average detection delay (reported in parenthesis) obtained using analytical formulas ('A1'), analytical formulas with skewness correction ('A2') and Monte Carlo thresholds (`MC') . The thresholds are obtained so that the ARL is set to be $2,000$. The amount of signal used correspond to the power setup described for Tables \ref{table:normal_general},  \ref{table:t5_var}, and \ref{table:ln_mean} in Section \ref{sec:Power}, respectively. It is clear that in the presence of change, the analytical formulas (with skewness correction) and Monte Carlo thresholds all lead to similar results in power and detection delay.

\begin{table}[ht]
\caption{Empirical ARL obtained from analytical $b_S$ such that $\boldE_\infty(T_S(b_S))=2,000$, $L=200$.}
\centering
\begin{tabular}{p{0.5cm}|c|c} 
\hline  
& $n_0=35$  & $n_0=40$ \\
\cline{1-3}
\multirow{1}{1.6cm}{$C1$} & 2286.56 & 2314.81\\
\hline
\multirow{1}{1.6cm}{$C2$} &  2330.53 & 2517.39\\
\hline
\multirow{1}{1.6cm}{$C3$} & 1855.41 & 2396.45\\
\hline 
\end{tabular}
\label{table:b_S_t5} 
\end{table} 

\begin{table}[ht]
\caption{Power and detection delay (reported in parenthesis) for $T_S(b_S)$ obtained using analytical thresholds (`A1') and Monte Carlo thresholds (`MC').}
\centering
\begin{tabular}{p{0.5cm}|cc|cc} 
\hline  
& \multicolumn{2}{c|}{$n_0 = 35$} &  \multicolumn{2}{c}{$n_0=40$} \\
\cline{1-5}
& A1 & MC & A1 & MC \\
\hline  
\multirow{2}{1.6cm}{$C1$} & 0.22 & 0.23  & 0.24 & 0.22  \\
& (67.72 & (67.46  & (66.71  & (70.72) \\
& $\pm 33.03)$ & $\pm 32.92)$ & $\pm 33.17)$ &  $\pm 34.88)$ \\
\hline
\multirow{2}{1.6cm}{$C2$} & 0.50 & 0.51 & 0.51 & 0.54\\
& (48.62 & (48.29 & (48.29 & (47.82 \\
& $\pm 19.01)$ & $\pm 19.00)$ & $\pm 18.99)$ & $\pm 19.35)$ \\
\hline
\multirow{2}{1.6cm}{$C3$} & 0.65 & 0.64 & 0.66 & 0.67\\
& (45.84 & (46.64 & (45.58 & (45.16 \\
& $\pm 24.71)$ & $ \pm 25.06)$ & $\pm 24.61)$ & $\pm 24.76)$ \\
\hline 
\end{tabular}
\label{table:b_S_t5_2} 
\end{table}

\begin{table}[ht]
\caption{Empirical ARL obtained from analytical $b_{W}$ such that $\boldE_\infty(T_{W}(b_{W}))=2,000$, $L=200$.}
\centering
\begin{tabular}{p{0.25cm}|cc|cc} 
\hline  
&\multicolumn{2}{c|}{$n_0 = 35$} &  \multicolumn{2}{c}{$n_0=40$} \\
\cline{1-5}
& A1 & A2 &  A1 & A2 \\
\hline 
\multirow{1}{1.2cm}{$C1$} & 1081.55 & 2216.99 & 1247.97 & 2363.42   \\
\hline
\multirow{1}{1.2cm}{$C2$} & 1430.04 & 2075.81 & 1619.72 & 2079.41 \\
\hline
\multirow{1}{1.2cm}{$C3$} & 1380.30 & 2086.30 & 1679.95 & 1904.12 \\
\hline 
\end{tabular}
\label{table:b_W_t5} 
\end{table} 

\begin{table}[ht]
\caption{Power and detection delay for  $T_W(b_W)$ obtained using analytical thresholds (`A1'), skewness corrected thresholds (`A2'), and Monte Carlo thresholds (`MC').}
\centering
\begin{tabular}{p{0.25cm}|p{0.9cm}p{0.9cm}p{0.9cm}|p{0.9cm}p{0.9cm}p{0.9cm}} 
\hline  
& \multicolumn{3}{c|}{$n_0 = 35$} &  \multicolumn{3}{c}{$n_0=40$} \\
\cline{1-7}
& A1 & A2 & MC & A1 & A2 &  MC \\
\hline 
\multirow{2}{1.2cm}{$C1$} & 0.21 & 0.13 & 0.13 & 0.21  & 0.13 & 0.12  \\
& (70.68 & (71.90 & (72.46 & (69.97 & (74.06 & (74.20 \\
& $\pm 39.21)$ & $\pm 35.87)$ & $\pm 36.03)$ & $\pm 39.51)$ & $\pm 36.77)$ & $\pm 38.05)$\\
\hline
\multirow{2}{1.2cm}{$C2$} & 0.48 & 0.46 & 0.46 & 0.49 & 0.47 & 0.46\\
& (46.25 & (47.18 & (47.85 & (45.77 & (47.01 & (47.13 \\
& $\pm 19.06)$ & $\pm 18.92)$ & $\pm 19.41) $ & $\pm 18.69)$ & $\pm 19.10)$ & $\pm 18.98)$ \\
\hline
\multirow{2}{1.2cm}{$C3$}  & 0.77 & 0.75 & 0.74 & 0.78 & 0.77 & 0.76\\
& (39.47 & (40.53 & (41.53 & (38.80 & (39.63 & (40.61\\
& $\pm 22.05)$ & $\pm 22.63)$ & $\pm 22.76)$ & $\pm 21.44)$ & $\pm 22.23)$ & $\pm 22.61)$ \\
\hline 
\end{tabular}
\label{table:b_W_t5_2} 
\end{table}

\begin{table}[ht]
\caption{Empirical ARL obtained from analytical $b_{M_\xi}$ such that $\boldE_\infty(T_{M_\xi}(b_{M_\xi}))=2,000$, $L=200$.}
\centering
\begin{tabular}{p{0.25cm}|cc|cc} 
\hline  
&  \multicolumn{2}{c|}{$n_0 = 35$} &  \multicolumn{2}{c}{$n_0=40$} \\
\cline{1-5}
&  A1 & A2 &  A1 & A2 \\
\hline 
\multirow{1}{1.2cm}{$C1$} & 1050.28  & 1980.73 & 1189.45 & 1997.47  \\
\hline
\multirow{1}{1.2cm}{$C2$} & 1390.37 & 1859.99 & 1507.60 & 2110.34 \\
\hline
\multirow{1}{1.2cm}{$C3$} & 1245.26 & 2053.76 & 1633.61 & 1935.36 \\
\hline 
\end{tabular}
\label{table:b_M_t5} 
\end{table} 

\begin{table}[ht]
\caption{Power and detection delay for $T_{M_\xi}(b_{M_\xi})$ obtained using analytical thresholds (`A1'), skewness corrected thresholds (`A2'), and Monte Carlo thresholds (`MC').}
\centering
\begin{tabular}{p{0.25cm}|p{0.9cm}p{0.9cm}p{0.9cm}|p{0.9cm}p{0.9cm}p{0.9cm}} 
\hline  
& \multicolumn{3}{c|}{$n_0 = 35$} &  \multicolumn{3}{c}{$n_0=40$} \\
\cline{1-7}
& A1 & A2 & MC & A1 & A2 &  MC \\
\hline 
\multirow{2}{1.2cm}{$C1$} & 0.28 & 0.19 & 0.17 & 0.26 & 0.21 &  0.22\\
& (66.55 & (72.48 & (72.51 & (68.67 & (70.89 & (70.04 \\
& $\pm 36.01)$ & $\pm 36.08)$ & $\pm 34.72)$ & $\pm 36.26)$  & $\pm 35.61)$ & $\pm 35.31)$ \\
\hline
\multirow{2}{1.2cm}{$C2$} & 0.61 & 0.55 & 0.50  & 0.63 & 0.58 & 0.56\\
&(46.49  & (47.68 & (48.27 & (46.32 & (46.83 & (47.54 \\
& $\pm 19.29)$ & $\pm 19.67)$ & $\pm 19.41)$ & $\pm 19.25) $ & $\pm 19.40)$ & $\pm 19.67)$ \\
\hline
\multirow{2}{1.2cm}{$C3$} & 0.75 & 0.73 & 0.71 & 0.75 & 0.75 & 0.74\\
& (40.76 & (41.83 & (43.11 & (40.46 & (41.41 & (41.86\\
& $\pm 22.67)$ & $\pm 22.87)$ & $\pm 23.10)$ & $\pm 22.54)$ & $\pm 23.01)$ & $\pm 22.86)$ \\
\hline 
\end{tabular}
\label{table:b_M_t5_2} 
\end{table} 

\section{Power assessment}\label{sec:Power} 

\subsection{Multivariate data} 

To examine the performance of the three new test statistics, 
we compare them to the existing approach in \cite{chen2019sequential} ($\max_t Z_{L|\by}(t,n)$) and two parametric likelihood-based approaches: Hotelling's $T^2$ test when there is change in mean and the generalized likelihood ratio test when there is variance change (both these two-sample tests are adapted to the scan statistic setting). The simulation setup is as follows: there are $N_0 = 200$ historical observations and a change occurs at $t=400$ ($200$ new observations after the start of the test). The observations are independent and follow a $d$-dimensional distribution. When there is a change in mean, the observations are shifted from $0$ by amount $\Delta$ in Euclidean distance. When the covariance matrix changes, to make the change less significant, only the first $d/5$ of the diagonal elements change with a multiple of $\sigma$, and the rest are unchanged. The amount of change is chosen so that the tests have moderate power to be comparable. For fair comparison, we use Monte Carlo simulations to determine the threshold for each of the test so that their average run lengths are all 2,000. 
Power is reported as the fraction of trials for which the change-point is detected within $50$ observations after the change occurred. In the following, we use `HT' to refer to the scan statistic over the Hotelling's $T^2$ statistic and use `GLR' to refer to the scan statistic over the generalized likelihood ratio statistic. For $d > 100$, in order for HT and GLR to be applicable in higher dimensions, we treat each data stream as if it is independent so that the covariance matrix's inverse and determinant are well-defined. 
 

When there is both mean and variance change, Table \ref{table:normal_general} shows the results under the Gaussian setting. When $d \le 500$, the GLR dominates in power. However, when the dimension increases, GLR is no longer able to retain competitive power compared to $S$ and $M$.  On the other hand, this setting is not well-suited for $W$ which is meant to capture mean change only and its performance is the worst here. 

To consider other distributions, we also compared the tests for multivariate log-normal data and multivariate $t_5$ data. The results for the log-normal data are shown in Table \ref{table:ln_mean}. 
Here there is a change in mean parameter only and $\Delta$ is chosen such that the location change dominates. In this setting, $W$'s performance dominates. We see that all the new tests outperform $Z$ and the parametric tests for $d >10$. 

The result for the multivariate $t_5$ data are shown in Table \ref{table:t5_var}. When there is a change in both mean and variance, $Z$ is unable to outperform the new test statistics. Among the new test statistics, their performance depends on whether the mean or variance signal dominates. When the mean change is stronger (for example, when $d=100$), $W$ performs comparably well. However, when the variance change is stronger (for example, when $d\ge1000$), $M$ and $S$ dominate.

\begin{table}[ht] 
\caption{Multivariate Gaussian data, mean and variance change.}
\begin{tabular}{p{0.4cm}|ccccc}
\hline
\hline
& \multicolumn{5}{c}{{Power}} \\
\hline
\hline
  d      & 10  & 100 &  500 & 1000 & 2000 \\
$\Delta$ & 0.35 & 0.5  & 0.9 & 1 & 0.85 \\
$\sigma$ & 0.55 & 0.65  & 0.8 & 0.9 & 0.9 \\
\hline
\hline
$HT$ & 0.02 & 0 & 0.003 & 0.008 & 0.005\\
 	 & (125.76) & - & (128.87) & (119.04) &  (120.64) \\ 
	 & $(\pm 45.92)$ & - & - & - & - \\ 
\hline
GLR & \textbf{0.34} & \textbf{1} & \textbf{1} &  0.17  & 0.42 \\
& \textbf{(59.81} & \textbf{(26.66} & \textbf{(29.99} & (73.21 & (57.13\\ 
& $\mathbf{\pm 23.77)}$ & $\mathbf{\pm 3.79)}$ & $\mathbf{\pm 4.23)}$ & $\pm 30.34)$ & $\pm 23.29)$ \\
\hline
$Z$   & 0.034 & 0.09 & 0.12 & 0.07 & 0.10 \\
& (101.70) & (89.82) & (81.23) & (92.61) & (83.69)\\ 
& $(\pm 32.32)$ & $(\pm 30.25)$  & $(\pm 28.08)$ &  $(\pm 33.36)$ & $(\pm 31.31)$ \\
\hline
$W$ & 0.13 & 0.17 &  0.14 & 0.08 & 0.08\\
& (72.46) & (60.34) & (57.20) & (65.67) & (62.81)\\ 
& $(\pm 36.03) $ & $(\pm 40.25)$ & $(\pm 36.58)$ & $(\pm 43.13)$ & $(\pm 42.16) $\\
\hline
$S$ & 0.23 & 0.83 &  0.98 & 0.78 & \textbf{0.98}\\
& (67.46) & (34.75) & (22.95) & (35.44) & \textbf{(22.22)} \\ 
& ($\pm 32.92)$ & $(\pm 20.00)$ & $(\pm 9.82)$ & $(\pm 19.88)$ & $\mathbf{(\pm 9.53)}$ \\
\hline
$M$ & 0.17 & 0.81 &  0.98 & \textbf{0.79} & \textbf{0.98} \\
& (72.51) & (35.07) & {(22.25)} & \textbf{(35.22)} & \textbf{(21.25)}\\ 
& $(\pm 34.72)$ & $(\pm 20.14) $ & $( \pm 10.23) $ & $\mathbf{(\pm 19.97)}$ & $\mathbf{(\pm 9.31)}$ \\
 \hline\hline
\end{tabular}
\label{table:normal_general} 
\end{table}


\begin{table}[ht]
\caption{Multivariate log-normal data, differ in the mean parameter.}
\begin{tabular}{p{0.4cm}|ccccc}
\hline
\hline
& \multicolumn{5}{c}{{Power}} \\
\hline
\hline
  d      & 10  & 100 &  500 & 1000 & 2000 \\
$\Delta$ & 0.95 & 1.6  & 1.9 & 2 & 2.1  \\
\hline
\hline
$HT$ &  \textbf{0.82} & 0.32 & 0.24 & 0.13 & 0.10 \\
  	& \textbf{(33.17} & (50.79 & (40.22 & (45.92 & (60.28 \\
	& $\mathbf{\pm 24.15)}$ & $\pm 35.58)$ & $\pm 17.60)$ & $\pm 29.19)$ & $\pm  40.27)$ \\
\hline
GLR &  0.06 & 0.09 &  0.07 &  0.05 &  0.03 \\
	& (93.59 & (60.75 & (71.51 & (78.72 & (68.55) \\
	& ${\pm 53.90)}$ & $\pm 43.58)$ & $\pm 56.15)$  & $\pm 55.72)$  & $\pm 57.57)$ \\
\hline
$Z$  & 0.43 & 0.37 & 0.25 & 0.21 & 0.17\\
	& (55.66 & (55.32 & (60.60 & (64.99 & (68.96 \\
	& ${\pm 29.73)}$ & $\pm  19.77)$ & $\pm 19.96)$ & $\pm 22.31)$ &$\pm 22.85)$   \\
\hline
$W$ &  0.43 & \textbf{0.86} & \textbf{0.84} & \textbf{0.74} & \textbf{0.62}\\
	& (56.46) & \textbf{(35.29)} & \textbf{36.67} & \textbf{(41.53)} & \textbf{(45.87)} \\
	& $\pm 31.97)$ & $\mathbf{\pm 15.97)}$ & $\mathbf{\pm 16.64)}$ & $\mathbf{\pm 22.76)}$ & $\mathbf{\pm 25.19)}$  \\
\hline
$S$ & 0.39 & 0.81 & 0.74 & 0.64 & 0.51 \\
	& (57.97 & (38.96 & (41.63 & (46.64 & (50.88 \\ 
	& $\pm 32.76)$ & $\pm 18.14)$ & $\pm  19.78)$ & $\pm 25.06)$ & $\pm 28.22)$    \\
\hline
$M$ & 0.41 & 0.85 & 0.81 & 0.71 & 0.59  \\	
	& (57.34) & (36.10) & (38.01) & (43.11) & (47.75) \\
	& $\pm 32.36)$ &  $\pm 16.49)$ & $\pm 17.32)$ & $\pm 23.10)$ & $\pm 27.25$) \\
 \hline\hline
\end{tabular}
\label{table:ln_mean} 
\end{table}




\begin{table}[ht] 
\caption{Multivariate $t$ data with $5$ degrees of freedom, mean and variance difference.}
\centering
\begin{tabular}{p{0.38cm}|cccccc}
\hline
\hline
& \multicolumn{5}{c}{{Power}} \\
\hline
\hline
  d      & 10   & 100 &  500 & 1000 & 2000 \\
$\Delta$ & 0.20 & 1.9  & 2.2 & 1.6 & 3.3  \\
$\sigma$ & 0.30 & 0.65  & 0.68 &  0.7 & 0.78 \\
\hline
\hline
$HT$  &  0.10   &  0   &  0.23  & 0.005    & 0.33 \\
 & (67.63) & - & (68.95) & (123.67) & (58.32) \\
 & $\pm 19.59)$ & - & $\pm 26.05)$ & -  & $\pm 21.57)$   \\
\hline
GLR  &  0.16   &   0.027 &  0.087   &  0.091    & 0.10 \\
 & (71.30 & (87.67 & (74.37 & (71.52 & (76.21 \\
 & $\pm 21.79)$ & $\pm 30.92)$ & $\pm 22.30)$ & $\pm 21.22)$  & $\pm 21.07)$  \\ 
\hline
$Z$  &  0.07   & 0.17    & 0.06     & 0.16     & 0.19 \\
 & (73.23)  & (73.05) & (74.67) & (70.31)  & (69.95)\\
 & $\pm 22.97)$ & $\pm 17.69)$ & $\pm 19.97)$ & $\pm 23.39)$  & $\pm 20.65)$  \\
\hline
$W$ &  0.10   & 0.46    & 0.36     & 0.13     & 0.34 \\
 &(55.32)  & \textbf{(47.85)}  & (44.94) & (44.41) & (44.17) \\
 & $\pm 22.91)$ & $\mathbf{\pm 19.41)}$  & $\pm 18.79)$  & $\pm 22.01)$  & $\pm 17.97)$ \\
\hline
$S$ 	&  \textbf{0.13}   & \textbf{0.51}   &  \textbf{0.53}    & 0.64     & \textbf{0.67}\\
 & \textbf{(58.31} & (48.29 & \textbf{(41.01} & (34.29 & \textbf{(36.79} \\
 & $\mathbf{\pm 23.06)}$ & $\pm 19.00)$ & $\mathbf{\pm 18.96)}$ & $\pm 16.74)$ & $\mathbf{\pm 15.80)}$  \\
\hline
$M$  &  0.09   & 0.50   &  0.47    & \textbf{0.67}     & 0.59\\
 & (58.32 & (48.27 & (42.97 & \textbf{(33.49} & (38.30 \\
 & $\pm 22.55)$ & $\pm 19.41)$ & $\pm 19.99)$ & $\mathbf{\pm 16.26)}$ & $\pm 17.47)$    \\
\hline\hline
\end{tabular}
\label{table:t5_var}
\end{table}
Based on the results of these tables, we see that the new graph-based methods perform well under various scenarios and have improved detection delay over the existing method in \cite{chen2019sequential}. In general, if one is certain that the change is locational, the test based on $W$ is recommended; while for more general changes, the tests based on $S$ and $M$ are recommended.

\subsection{Network data} 
Non-Euclidean data, also referred to as object data, is simply data that does not lie in the Euclidean space. Examples include networks, images, shapes, and trees. Fundamental statistical tools involving vector space analysis are no longer applicable for object data, making it challenging to directly analyze these data types. To demonstrate the new test statistics' power on non-Euclidean data, we evaluate the proposed methods on a sequence of networks. We generate random networks using the configuration model, which is a specific method that allows us to generate random networks with a specified degree sequence (please see \cite{molloy1995critical, newman2018networks} for an overview). For every node with degree $k_i$, we create $k$ half-edges (referred to as `stubs'). The network is created by iteratively selecting two stubs uniformly at random and connecting them to form an edge. This is done until no stubs remain and will result in a network with a pre-defined degree sequence. For the graph-based results, we show the results for graphs constructed using two different similarity measures. Specifically, for a network at time $t$, we encode the network using an adjacency matrix $A_t$ with $1$ for element $(i,j)$ if node $i$ and $j$ are connected, and 0 otherwise. The similarity measures are: 
\begin{enumerate} 
\item Similarity 1: $ ||A_t - A_s ||_F$ ,
\item Similarity 2: $\frac{||A_t - A_s ||_F}{\sqrt{||A_t||_F* ||A_s||_F}}.$
\end{enumerate} 

Under this setting, we compare the graph-based approach to another method designed specifically to detect changes in a stream of networks/graphs \cite{zambon2018concept}. The approach proposed in \cite{zambon2018concept} is quite general: it does not make explicit assumptions on the network/graph and can be applied to a stream of graphs of varying sizes.  Their approach consists of two steps: (1) each graph $G_t$ is mapped to a vector $y_t$ through a prototype-based embedding, and (2) a change is then detected in the stream of vectors $y_t$ using any conventional multivariate change-detection procedure. Specifically, embedding is carried out by assessing the dissimilarity between graphs and a selection of prototypes. In the paper, they rely on the graph edit distance (GED), which count and weights the edit operations that are needed in order to make two input graphs equal and is applicable to graphs where the nodes are unidentified.  To carry out their approach, we follow the implementation proposed in \cite{zambon2018concept}; specific details can be found in Section IV of \cite{zambon2018concept}.

Since computing the GED is quite computationally expensive, we allow each network to have only 6 nodes. We set a change at $\tau=101$. Before and after the change, the total node degree remains the same, but the sequence of node degree changes. Before the change, each network is constructed such that nodes 1 and 4 have node degree of 1 and the remaining nodes have node degree 2. After the change, nodes 3 and 6 have node degree of 1 and the remaining nodes have node degree 2. Observe that in this setting Similarity 1 and 2 are equivalent and so we only report results for Similarity 1. The first $50$ observations are treated as training observations for \cite{zambon2018concept}. We set $L = 50$ for the graph-based approach. For fair comparison with the method in \cite{zambon2018concept},  we set the window size to be 50 and we implement their method using both GED and the Forbenius norm (Similarity 1) as the similarity measure between networks. The ARL is set to be 2,000. Power is defined as the number of trials (out of 100) where the change is detected within 50 observations. Among those trials where the change is detected, we also report the expected detection delay (EDD) and its standard deviation when applicable. We note that the approach in \cite{zambon2018concept} depends on the choice of embedding and its hyper-parameters, which should be chosen carefully. A thorough discussion is provided in \cite{zambon2018concept}. We carry out simulations under a variety of settings for the number of dimensions to embed ($d$) and the number of prototypes to select (nproto).

Table \ref{table:networkCompare} show the power performance of the graph-based methods compared to the approach in \cite{zambon2018concept}. We can see that all of the graph-based statistics do better in terms of power and detection delay compared to \cite{zambon2018concept}. 

\begin{table}[ht]
\caption{Power comparison for configuration network model.}
\centering
\begin{tabular}{|p{4.5cm}|cc|}
\hline
& Power&  EDD \\
\hline
Zambon et. al \cite{zambon2018concept} (d=5,  nproto = 5) & & \\
\hspace{2mm} GED  & 2 & 51.00 $\pm$ NA  \\
\hspace{2mm} Similarity 1  & 4 & $26 \pm 25$ \\
\hline
Zambon et. al \cite{zambon2018concept}  (d=10,  nproto = 10) & & \\
\hspace{2mm} GED & 12 & 17.67 $\pm$ 23.57  \\
\hspace{2mm} Similarity 1  & 4 & $26 \pm 25$ \\
\hline
Zambon et. al \cite{zambon2018concept} (d=15,  nproto = 15) & & \\
\hspace{2mm} GED  & 26 & 24.08 $\pm$ 24.93  \\
\hspace{2mm} Similarity 1  & 20 & $46 \pm 15$ \\
\hline
$Z$ && \\
\hspace{2mm} Similarity 1 & 92 & $20.12 \pm 5.03$   \\
\hline
$W$  && \\
\hspace{2mm}  Similarity 1 & 92 & $21.45 \pm 5.28$  \\
\hline
$S$ 	&& \\
\hspace{2mm} Similarity 1 & 82 & $25.89 \pm 7.16$ \\
\hline
$M$ && \\
\hspace{2mm} Similarity 1 & 82 & $21.09 \pm 7.42$ \\
\hline
\end{tabular}
\label{table:networkCompare}
\end{table}

For further exploration of the graph-based approach, we generate a sequence  of networks such that each network has $20$ nodes with a pre-specified degree sequence (see below for details). A change in the sequence of networks happens at $\tau = 101$. The first $50$ networks in the sequence are treated as historical observations. Power is reported as the fraction of trials for which the change-point is detected within 50 observations after the change occurred. Among those trials where the change is detected within the first 50 observations, the expected detection delay and its standard deviation is also reported (EDD). 

The networks are generated under two different settings: 
\begin{enumerate} 
\item A fixed degree change in the network: before the change, half of the nodes have out-degree and in-degree 1 and half the nodes have out-degree and in-degree 3. After the change, 5 nodes have out-degree and in-degree $k_1$ and 5 nodes have out-degree and in-degree $k_2$. The remaining nodes remain unchanged.
\item A random degree change in the network: before the change,  half of the nodes have out-degree and in-degree 1 and half the nodes have out-degree and in-degree 3. After the change,  half of the nodes have out-degree and in-degree randomly selected from $k_3$ to $k_4$ and the remaining half of the nodes have out-degree and in-degree 3.
\end{enumerate} 

Tables \ref{table:network_power1} and \ref{table:network_power2} report the power and expected detection delay of the graph-based test statistics for similarity measure 1 and 2, respectively. For similarity measure 1, the performance of $S$ and $M$ dominate in almost all settings, with the exception of when $k_1 =2 , k_2 = 2$ ; in general $S$ and $M$ do well with respect to both power and expected detection delay. For similarity measure 2, the performance of $W$ and $Z$ improve substantially, while the performance of $S$ and $M$ remain stable.

\begin{table}[ht]
\caption{Power comparison of graph-based methods for configuration network model with 20 nodes under similarity measure 1.}
\centering
\begin{tabular}{p{0.3cm}|cc|cc}
\hline
\hline
& \multicolumn{4}{c}{Fixed degree} \\
\hline
&\multicolumn{2}{c|}{ $k_1 = 2$, $k_2 =2$} & \multicolumn{2}{c}{ $k_1 = 4$, $k_2 = 5$}\\
& Power&  EDD & Power & EDD  \\
\hline
\hline
$Z$  	 & 74 &  $16.32 \pm 4.40$  & 14 & $23.71 \pm 8.00$   \\
\hline
$W$  & 80 & $25.41 \pm 6.58$  & 57 & $36.42 \pm 4.32$  \\
\hline
$S$ 	 &72& $10.65 \pm 3.42$ & 71 & $8.97 \pm 2.50$ \\
\hline
$M$  & 60 &  $9.08 \pm 4.06$ & 55 & $7.94 \pm 2.38$ \\
\hline\hline
\end{tabular}
\vspace*{0.5cm}

\begin{tabular}{p{0.3cm}|cc|cc}
\hline
\hline
& \multicolumn{4}{c}{Random degree} \\
\hline
& \multicolumn{2}{c|} {$k_3 = 1$, $k_4 = 3$} & \multicolumn{2}{c} {$k_3 = 1$, $k_4 = 7$}\\
& Power&  EDD & Power & EDD \\
\hline
\hline
$Z$  & 0 & NA &  41 & $18.26 \pm 5.83$ \\
\hline
$W$ &  1 & $45 \pm -$ & 55 & $31.41 \pm 4.40$ \\
\hline
$S$ 	 & 78 & $13.42 \pm 3.12$ & 65 & $9.32 \pm 2.72$\\
\hline
$M$  &  59 & $12.05 \pm 2.98$ & 48 & $8.83 \pm 3.03$ \\
\hline\hline
\end{tabular}
\label{table:network_power1}
\end{table}

\begin{table}[ht]
\caption{Power comparison of graph-based methods for configuration network model with 20 nodes under similarity measure 2.}
\centering
\begin{tabular}{p{1cm}|cc|cc}
\hline
\hline
& \multicolumn{4}{c}{Fixed degree} \\
\hline
&\multicolumn{2}{c|}{ $k_1 = 2$, $k_2 =2$} & \multicolumn{2}{c}{ $k_1 = 4$, $k_2 = 5$} \\
& Power&  EDD & Power & EDD  \\
\hline
\hline
$Z$  	 & 74 & $16.32 \pm 4.40$  & 2 & $42.5 \pm -$  \\
\hline
$W$  & 80 & $25.41 \pm 6.58$ & 4 & $25.75 \pm -$  \\
\hline
$S$ 	 &72& $10.65 \pm 3.42$ & 65 & $7.66 \pm 1.06$ \\
\hline
$M$  & 60 & $9.08 \pm 4.06$ & 51 & $7.98 \pm 2.45$ \\
\hline\hline
\end{tabular}
\vspace*{0.5cm}

\begin{tabular}{p{1cm}|cc|cc}
\hline
\hline
& \multicolumn{4}{c}{Random degree} \\
\hline
& \multicolumn{2}{c|} {$k_3 = 1$, $k_4 = 3$} & \multicolumn{2}{c} {$k_3 = 1$, $k_4 = 7$}\\
& Power&  EDD & Power & EDD \\
\hline
\hline
$Z$  & 23 & $22 \pm 8.10$ &  1 & $46 \pm -$ \\
\hline
$W$ &  64 & $13.38 \pm 3.39$ & 12 & $13.30 \pm 5.69$ \\
\hline
$S$ 	 & 65 & $10.45 \pm 2.07$ & 63 & $9.15 \pm 2.75$\\
\hline
$M$  &  58 & $12.03 \pm 2.73$ & 48 & $12.17 \pm 2.39$ \\
\hline\hline
\end{tabular}
\label{table:network_power2}
\end{table}

\section{A real data application} \label{sec:Application} 
We compare the new approaches to the method in \cite{chen2019sequential} using the  yellow taxi trip records data. The data set is publicly available on the NYC Taxi \& Limousine Commission (TLC) website. It provides information on the taxi pickup and drop-off date/times, longitude and latitude coordinates of pickup and drop-off locations, trip distances, fares, rate types, payments types, and driver-reported passenger counts. 

Based on this data set, a natural question to ask is: Can we detect a change in traffic patterns during peak travel seasons? Here, we focus on those trips that began at John F. Kennedy International Airport and we look at two different time periods: the months of June through August and November through December in 2015. The dataset has been completely collected at the time of analysis. However, we treat it as if the data were being observed in order to illustrate how the proposed method works. For simplicity, the boundary of JFK airport was set to be $40.63$ to $40.66$ latitude and $-73.80$ to $-73.77$ longitude. 

For those trips that began with a pickup at JFK, we extract information on their longitude and latitude drop-off coordinates. Using longitude/latitude coordinates, we create a $30$ by $30$ grid of New York City and count the number of taxi drop-offs that fall within each cell, where each cell represents a longitude, latitude coordinate range. Then for each day, we have a $30$ by $30$ matrix such that each element represents the number of taxi drop-offs in each location. 

\begin{table}[ht]
\caption{Detected stopping times for NYC taxi pickups from JFK for June 1, 2015 - August 31, 2015.}  \label{table:pickup2}
\centering
\begin{tabular}{|c|c|c|} 
 
\hline
 & Reported stopping times & Estimated change-point \\
\hline
\multirow{1}{*}{$Z$}  & --- & ---  \\
\hline
\multirow{ 1}{*}{$W$}  & 07/03 - 07/04 & 06/29 (Day 30) \\
\hline
\multirow{1}{*}{$M$} & 07/03 - 07/04 & 06/29 (Day 30)\\
\hline
\multirow{ 1}{*}{$S$}  & 07/03 - 07/05 & 06/29 (Day 30) \\
\hline
\end{tabular}
\end{table}

\begin{table}
\caption{Detected stopping times for NYC taxi pickups from JFK for October 21, 2015 - December 31, 2015.}  \label{table:pickup1}
\centering
\begin{tabular}{|p{0.5cm}|c|c|c|} 
 
\hline
 & Reported stopping times & Estimated change-point \\
\hline
\multirow{ 3}{*}{$Z$}  &  11/27 - 11/31 & 11/21 (Day 32) \\
& 12/23 - 12/25 & 12/10 (Day 51)\\
 &  12/30 - 12/31  & 12/26 (Day 67) \\
\hline
\multirow{3}{*}{$W$}  & 11/28  & 11/21 (Day 32) \\
& 12/23 - 12/26  & 12/19 (Day 60) \\
 &  12/29 - 12/31  & 12/26 (Day 67) \\
\hline
\multirow{3}{*}{$M$} & 11/28 & 11/21 (Day 32) \\
& 12/23 - 12/26  & 12/19 (Day 60) \\
&  12/29 - 12/31  & 12/26 (Day 67) \\
\hline
\multirow{ 3}{*}{$S$}  & 11/27 - 11/30 & 11/21 (Day 32), 11/23 (Day 34) \\
&  12/23 - 12/26 & 12/19 (Day 60) \\
&  12/29 - 12/31  & 12/26 (Day 67) \\
\hline
\end{tabular}
\end{table}


\begin{figure}
    \centering
    \hspace{-6mm}
    \subfloat{
       \includegraphics[width=0.24\textwidth]{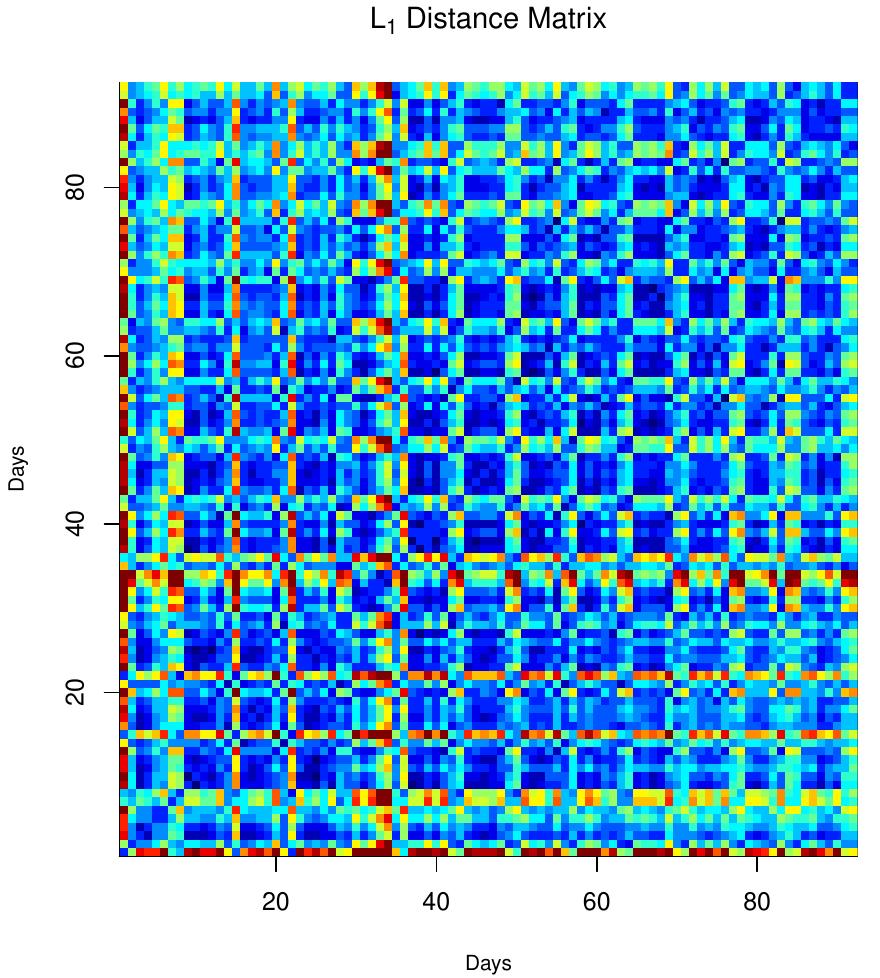}
       }
       \hspace{-4mm}
          \subfloat{
       \includegraphics[width=0.24\textwidth]{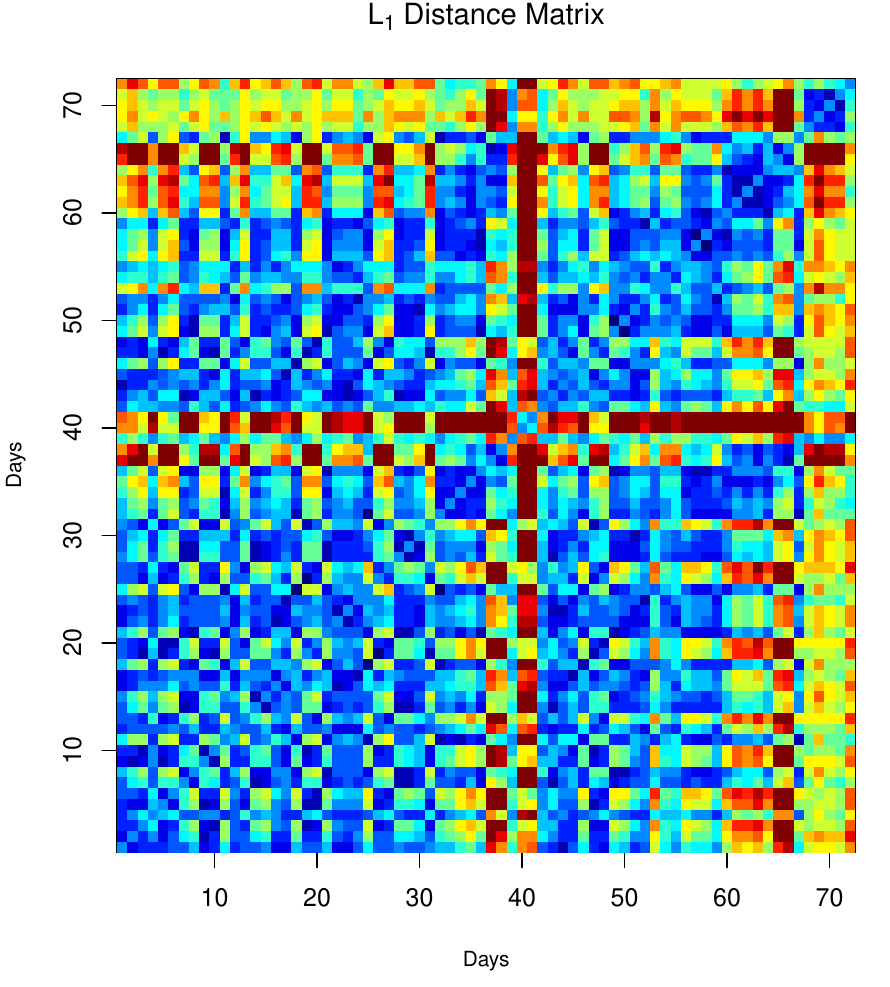}
       }
        \caption{Left panel: Heatmap of $L_1$ norm distance matrix of vector $v_i$ for $i = 1, \hdots 93$, corresponding to dates June 1, 2015 - Aug. 31, 2015. Right panel: Heatmap of $L_1$ norm distance matrix of vector $v_i$ for $i = 1, \hdots 72$, corresponding to dates Oct. 21, 2015 - Dec. 31, 2015.}
    \label{fig:distance_heatmap}
 \end{figure}
We apply the new approaches, as well as $Z$, to detect changes in the months of June through August 2015. We use data from the month of May as historical data. Applying the offline change-point detection method in \cite{chen2015graph} and \cite{chu2019asymptotic} on the observations in May, we find there is no change-point in the first 30 days, so we set $L=30$, $n_0  = 5$, and $n_1 = L - n_0$. We denote $A_i$ to be the $30$ by $30$ matrix on day $i$ and $v_i$ to be the vector form of $A_i$, which is now $900$ by $1$. The $L_1$ norm is used to construct the $k$-NN graph representing similarity between days. Here, the new test statistics ($W$, $M$, and $S$) all report a stopping time of July 3 and July 4 whereas $Z$ is unable to detect any anomaly event (Table \ref{table:pickup2}). The change-point triggering these stopping times is estimated to be June 29. To perform a sanity check, we plot a heatmap of the $L_1$ distance matrix used to the construct the $k$-NN graph (see Figure \ref{fig:distance_heatmap}, left panel). Based on the heatmap, we can see there is a clear signal happening around Day 30, which corresponds with the results from the new test statistics. 

To detect changes in November and December 2015, we use data from the months of September and October 2015 as historical data. Applying the offline change-point detection method in \cite{chen2015graph} and \cite{chu2019asymptotic} on the observations in September and October, we find there is no change-point in the first 50 days. Therefore, we treat the first 50 observations from Sept. 1 - Oct. 20 as historical observations and we begin the test at Oct. 21. We set $L = 50$, $n_0 = 8$, and $n_1 = L - n_0$. The stopping times based on the new test statistics report back dates that seem to be quite reasonable (see Table \ref{table:pickup1}). We see that multiple stopping times are caused by the same anomaly event. When the signal is large enough, the new test statistics ($W$, $M$, and $S$) and $Z$ perform similarly: all are able to detect a change in travel pattern close to Thanksgiving and the Christmas holidays.  Again to check our results, we plot a heatmap of the $L_1$ distance matrix used to the construct the $k$-NN graph. We can see that there is a clear signal starting roughly around Day 30 and again around Day 60 and Day 67, which matches the results reported from the test statistics. In comparison with the heatmap from the months of June through August, the signal from the summer months is much weaker and in that case $Z$ is unable to detect any anomaly event.

\section{Conclusion} \label{sec:Disc}
We propose new graph-based test statistics under $k$-NN for detecting change-points sequentially as data are generated. We study the asymptotic properties of the stopping rules based on the new test statistics, and derived the analytic formulas to approximate the average run lengths of the new stopping rules.  To accommodate finite samples,  skewness corrected approximations were also derived for the weighted and max-type edge-count statistic under $k$-NN. The skewness-corrected versions give much more accurate approximations to the average run lengths and can be used reliably in practice.  The performance of the proposed test statistics are examined under various common scenarios. Simulation studies reveal that the new test statistics have shorter detection delays for a wider range of alternatives and exhibit  power gains for scale change when compared to parametric tests and the test statistic proposed in \cite{chen2019sequential}. Specifically, simulation results show that the weighted-edge count statistic ($W$) is useful at quickly detecting mean changes. When a change in variance is also of interest, the generalized edge-count statistic ($S$) and max-type edge-count statistic ($M$) are more effective in detecting changes and obtain faster detection. Together with the fact that skewness corrected average run length approximations can be obtained for the max-type edge-count statistic, the stopping rule $T_{M_\xi}$ is recommended for sequential detection of general changes.

\section*{Acknowledgments}
Hao Chen was supported in part by NSF Grants DMS-1513653 and 1848579.
%

\bibliographystyle{IEEEtran}
\bibliography{references_online}

%
%
%

\end{document}